\documentclass{elsart}
\usepackage[applemac]{inputenc}
\usepackage{graphicx}
\usepackage{amsmath}
\usepackage{times}
\usepackage{txfonts}
\usepackage{array}
\usepackage{natbib}
\usepackage{rotating}
\usepackage[dvipsnames,usenames]{color}
\usepackage{setspace}
\usepackage{setspace}

\newcommand{\parallelsum}{\mathbin{\!/\mkern-4mu/\!}}
\newcommand{\te}[1]{10^{#1}}
\newcommand{\ut}[1]{\hspace{1mm}\mathrm{#1}}
\newcommand{\vc}[1]{\boldsymbol{\mathrm #1}}
\newcommand{\vecu}{\vc{u}} 
\newcommand{\vecB}{\vc{B}} 
\newcommand{\vecom}{\vc{\omega}}

\newcommand{\cit}[1]{\let\temp=\\ \centering #1\let\\=\temp}
\newcommand{\rev}[1]{{#1}}

\newcolumntype{L}[1]{>{\raggedright\let\newline\\\arraybackslash\hspace{0pt}}m{#1}}
\newcolumntype{C}[1]{>{\centering\let\newline\\\arraybackslash\hspace{0pt}}m{#1}}
\newcolumntype{R}[1]{>{\raggedleft\let\newline\\\arraybackslash\hspace{0pt}}m{#1}}

\begin{document}

\begin{frontmatter}
\title{Approaching Earth’s core conditions in high-resolution geodynamo simulations}
\author{Julien Aubert}
\address{Université de Paris, Institut de physique du globe de Paris, CNRS, F-75005 Paris, France.}

\begin{abstract}
The geodynamo features a broad separation between the large scale at which Earth's magnetic field is sustained against ohmic dissipation and the small scales of the turbulent and electrically conducting underlying fluid flow in the outer core. Here, the properties of this scale separation are analysed using high-resolution numerical simulations that approach closer to Earth's core conditions than earlier models. The new simulations are obtained by increasing the resolution and gradually relaxing the hyperdiffusive approximation of previously published low-resolution cases. This upsizing process does not perturb the previously obtained large-scale, leading-order quasi-geostrophic (QG), and first-order magneto-Archimedes-Coriolis (MAC) force balances. As a result, upsizing causes only weak transients typically lasting a fraction of a convective overturn time, thereby demonstrating the efficiency of this approach to reach extreme conditions at reduced computational cost. As Earth's core conditions are approached in the upsized simulations, Ohmic losses dissipate up to 97 percent of the injected convective power. Kinetic energy spectra feature a gradually broadening self-similar, power-law spectral range extending over more than a decade in length scale. In this range, the spectral energy density profile of vorticity is shown to be approximately flat between the large scale at which the magnetic field draws its energy from convection through the QG-MAC force balance and the small scale at which this energy is dissipated. The resulting velocity and density anomaly planforms in the physical space consist in large-scale columnar sheets and plumes, respectively co-existing with small-scale vorticity filaments and density anomaly ramifications. In contrast, magnetic field planforms keep their large-scale structure after upsizing. The small-scale vorticity filaments are aligned with the large-scale magnetic field lines, thereby minimising the dynamical influence of the Lorentz force. The diagnostic outputs of the upsized simulations are more consistent with the asymptotic QG-MAC theory than those of the low-resolution cases that they originate from, but still feature small residual deviations that may call for further theoretical refinements to account for the structuring constraints of the magnetic field on the flow.
\end{abstract}

\begin{keyword}
Dynamo: theories and simulations; Core; Numerical modelling.
\end{keyword}

\end{frontmatter}

\section{\label{intro}Introduction}
Steadily improving numerical models of the geodynamo have appeared since the historical outset of the discipline, more than two decades ago. In retrospect, it is quite striking that a significant part of the fundamental insight on how Earth's magnetic field may be regenerated by the underlying core convection has been gained through early models that featured a low spatial resolution. A self-sustained magnetic field with dipole reversals \citep{Glatzmaier1995} and basic but plausible induction mechanisms \citep{Olson1999} could indeed be captured with simulations resolving spatial scales down to only a thousand kilometers at Earth's core-mantle boundary (corresponding to spherical harmonic degree 20). At first glance this may seem sufficient since geomagnetic field models using even the most recent satellite observations can resolve the main field of internal origin only up to degree 13 and its rate of change up to degree about 16, after which 
contributions from other sources and contamination from noise become too strong \citep{Hulot2015}. This is of course misleading because the Earth's core is expected to be in a strongly turbulent state \citep[e.g.][]{Aubert2017}, implying that the large-scale picture that we observe should be influenced by non-linear interactions between much smaller, unobservable scales. This theoretical expectation has been partly confirmed by subsequent modelling efforts, where advances of the computer power allowed for calculations with a gradually increasing spatial resolution (harmonic degree in excess of 100), and more realistic physical parameters. The output from these models has indeed reached a higher level of quantitative agreement with the detailed morphology of the geomagnetic field \citep{Christensen2010, Mound2015}, and of its temporal variations, notably the westward drift of low-latitude magnetic flux patches at the core-mantle boundary \citep{Aubert2013b,Schaeffer2017}. However, this new generation of models still featured a relatively low spatial resolution and parameters quite far from Earth's core conditions, raising the question of whether these successes were indeed obtained in physically relevant conditions, especially regarding the force balance obtained in the numerical models \citep[e.g.][]{Soderlund2012}. 

In an attempt to answer this question, it has recently been shown that the physical conditions at which these models operate and those of Earth's core can be connected through the formulation of a unidimensional theoretical path in model parameter space \citep{Aubert2017}. The level of magnetic turbulence (as characterised by the value of the magnetic Reynolds number, see section \ref{model}) is already realistic at path start and therefore remains constant along this path. However, a decreasing ratio between the kinematic and magnetic diffusivities along the path (decreasing magnetic Prandtl number) and increasing levels of forcing imply that hydrodynamic turbulence becomes stronger (increasing Reynolds number), and separation between the largest magnetic field scales and the smallest velocity field scales increases accordingly. Despite this increasing level of turbulence, a striking result of calculations carried out along this path is that at large-scales (below spherical harmonic degree 30), and time scales longer than the convective overturn time, the morphology of the magnetic field, of its rate-of-change and underlying core flows, as well as the leading-order force balance remain approximately invariant as we progress towards Earth's core conditions \citep{Aubert2017,Aubert2018}. This result rationalises the early successes of geodynamo modelling that were obtained at low spatial resolution, but raises further questions regarding the role and importance of the broad scale separation that is expected in Earth's core, and indeed observed in recent landmark simulations at extreme conditions and high resolution \citep[harmonic degrees in excess of 200 and up to 1000,][]{Sakuraba2009,Yadav2016PNAS,Schaeffer2017,Sheyko2018}. One key point is that only a small fraction (the first 15 percent) of this parameter space path has been covered by fully resolved direct numerical simulations (DNS), and a larger portion (up to 50 percent) has been explored using lower-resolution, large-eddy simulations (LES) employing a form of eddy viscosity that prescribes the smallest possible scales in the system. For these simulations, hyperdiffusion has been preferred over sophisticated subgrid parameterisations \citep[e.g.][]{Baerenzung2008,Baerenzung2010,Matsui2013}, \rev{on the basis of} physical scaling arguments \citep{Davidson2013} that predict that the dominant force balance in the system is achieved at a relatively large scale (spherical harmonic degree about 10) that does not significantly evolve along the path. The realism of LES simulations has been successfully tested in the path region where both DNS and LES are feasible \citep{Aubert2017}. However, this region corresponds only to a moderate level of hydrodynamic turbulence, and one may question the realism of LES as regards the details of the scale separation and the associated non-linear energy transfers when turbulence significantly increases. This calls for high-resolution geodynamo simulations at advanced positions along this path.

High-resolution geodynamo simulations in the rapidly rotating and strongly turbulent regime pertaining to Earth's core conditions are also desirable to advance the understanding of how rotation and a self-generated magnetic field constrain turbulence in such systems. Recent cross-fertilisation between theoretical investigations \citep{Davidson2013, Calkins2015, Aurnou2017, Calkins2018} and numerical efforts \citep{Aubert2017, Schaeffer2017} have led to an emerging consensus that in this regime, the system is governed at large scales by a leading-order geostrophic balance between the pressure and Coriolis forces, as opposed to the earlier conjecture of a large-scale magnetostrophic balance where the Lorentz force would also reach a leading order \citep[e.g.][]{Hollerbach1996}. With a leading-order geostrophic balance being enforced, it has been argued that the planforms of the dynamo are in fact broadly similar to those of non-magnetic, rotating convection \citep{Soderlund2012}. This apparently weak influence of the magnetic field was however obtained in contexts where the magnetic energy was equivalent to the kinetic energy, implying a weak magnetic control on the flow. Subsequent simulations at more extreme parameters and higher magnetic to kinetic energy ratios (stronger magnetic control) have indeed highlighted different distributions of length scales in  magnetic and non-magnetic rotating turbulence \citep{Yadav2016PNAS,Sheyko2018}. Still, the Lorentz force tends to be minimised as a consequence of Lenz law, leading for instance to the enforcement of the Taylor constraint \citep{Taylor1963} at the axisymmetric level \citep[see e.g.][]{Aubert2017}. Though there exists an important body of numerical work on rotating or magnetic turbulence in cartesian domains \citep[see reviews in][]{Tobias2012, Nataf2015}, the configuration of forced magnetohydrodynamic and rotating turbulence in a spherical shell has been less studied in this so-called strong regime of interaction between the velocity and magnetic fields. 

In the present study, the hyperdiffusive approximation of LES simulations previously published in \cite{Aubert2017,Aubert2018} is gradually relaxed, and their spatial resolution is increased accordingly, \rev{with all other parameters kept the same}, a process that will be referred to as upsizing. Due to computational limitations, high spatial resolution and long integration times cannot be achieved simultaneously. The upsized models are therefore computed only for a short amount of time and our first goal is to assess the duration of transients caused by upsizing and the relevance of this numerical approach. Our next goals are to evaluate the successes and shortcomings of the previously employed hyperdiffusive approximations in the light of the upsized solutions and of other available extreme DNS simulations, and propose optimal approximations for forthcoming work. The upsized simulations are also useful to evaluate the robustness of the leading-order force balance, extract the next-order force balances established at smaller scales, characterise the turbulence and scale separation through their statistical properties and assess their level in Earth's core. The manuscript is organised as follows: section \ref{model} presents the numerical dynamo model and methods. Results are presented in section \ref{results} and discussed in section \ref{discu}. 

\section{\label{model}Model and methods}
\subsection{Model set-up, dimensionless inputs and outputs}
We implement the equations of Boussinesq convection, thermochemical density anomaly transport and magnetic induction in the magnetohydrodynamic approximation within an electrically conducting and rotating spherical fluid shell of thickness $D=r_{o}-r_{i}$ and aspect ratio $r_{i}/r_{o}=0.35$ representing the Earth's outer core. These equations can be found in  \cite{Aubert2017} (from hereafter A17). We solve for the velocity field $\vecu$, magnetic field $\vecB$ and density anomaly field $C$. The shell is surrounded by solid inner core of radius $r_{i}$, and a solid mantle between radii $r_{o}$ and $1.83 r_{o}$, both of which are electrically conducting. These two surrounding layers are electromagnetically coupled to the outer core and gravitationally coupled between each other, and present a variable axial differential rotation with respect to the fluid shell. The moments of inertia of the three regions are set in Earth-like proportions, and the constant angular momentum of the ensemble defines the planetary rotation rate vector $\vc{\Omega}$. Convection is driven from below using a homogeneous mass anomaly flux $F$ imposed at radius $r_{i}$, and a volumetric buoyancy sink term such that there is no homogeneous mass anomaly flux at radius $r_{o}$. The mechanical, thermochemical and electromagnetic boundary conditions are respectively of the stress-free, fixed-flux and electromagnetically conducting types at both boundaries. The complete set-up of boundary conditions, lateral heterogeneities and core-mantle-inner core couplings corresponds to the Coupled Earth model described in \cite{Aubert2013b}; A17; \cite{Aubert2018}, and full details on this set-up can be found in these earlier studies. All models produced a self-sustained and dipole-dominated magnetic field with Earth-like geometry \citep[][A17]{Christensen2010} that did not reverse polarity during their integration time. 

The main model parameters are the flux-based Rayleigh, Ekman, Prandtl and magnetic Prandtl numbers
\begin{eqnarray}
Ra_{F}&=&\dfrac{g_{o}F}{4\pi\rho\Omega^{3}D^{4}},\\
E&=&\dfrac{\nu}{\Omega D^{2}},\\
Pr&=&\dfrac{\nu}{\kappa},\\
Pm&=&\dfrac{\nu}{\eta}.
\end{eqnarray}
Here $g_{o}$, $\rho$, $\nu$, $\kappa$ and $\eta$ are respectively the gravity at radius $r_{o}$, the fluid density, viscosity, thermo-chemical and magnetic diffusivities. We use the path theory (A17) that bridges the parameter space gap between our previous coupled Earth model \citep{Aubert2013b} and Earth's core conditions by relating these four parameters to a single variable $\epsilon$:
\begin{eqnarray}
Ra_{F}&=&\epsilon Ra_{F} (\mathrm{CE}),\\
E&=&\epsilon E (\mathrm{CE}),\\
Pr&=&1,\\
Pm&=&\sqrt{\epsilon} Pm (\mathrm{CE}).
\end{eqnarray}
Here $Ra_{F}(CE)=2.7~\te{-5}$, $E(CE)=3~\te{-5}$ and $Pm(CE)=2.5$ are the control parameters of the coupled Earth dynamo model. Through these definitions, A17 introduced the concept of a unidimensional path in parameter space, starting at $\epsilon=1$, and ending at $\epsilon=\te{-7}$ where reasonable estimates for Earth's core conditions are obtained (A17). The model cases presented here (Table \ref{bigtable1}) range from $\epsilon=0.1$ to $\epsilon=3.33~\te{-4}$, this latter value defining the model at the middle of the path, which to date is the closest to Earth's core conditions computed in a numerical simulation. The numerical implementation involves a decomposition of $\vecu$, $\vecB$ and $C$ in spherical harmonics up to degree and order $\ell_{\mathrm{max}}$ and a discretisation in the radial direction on a second-order finite-differencing scheme with $NR$ grid points. We use the spherical harmonics transform library  SHTns \citep{Schaeffer2013} freely available at {\tt https://bitbucket.org} {\tt/nschaeff}{\tt/shtns}. Full details on the numerical implementation can be found in A17. 

Along the parameter space path, the magnetic Reynolds number $Rm=U D/\eta$ (where $U$ is the r.m.s value of $\vc{u}$ in the shell) remains approximately constant and Earth-like (Table \ref{bigtable2}), a defining property of this path. As a consequence, the level of hydrodynamic turbulence increases as we progress towards Earth's core conditions, as witnessed by the increase of the hydrodynamic Reynolds number $Re=U D / \nu=Rm/Pm$ reaching values up to $Re\approx 24000$. \rev{A17 explored a parameter range down to $\epsilon=0.1$ (the first 14 percent of the path on a logarithmic scale) with DNS simulations.} Beyond this point, DNS simulations become computationally very expensive \citep{Schaeffer2017}. Therefore, in A17 the remaining portion to the middle of the path was explored with LES simulations where a hyperdiffusive approximation was used on the velocity and density anomaly fields, but not on the magnetic field which remained fully resolved. The principle is to use values of $\ell_{\mathrm{max}}$ and $NR$ lower than those required by DNS, and effective diffusivities $\nu_\text{eff},\kappa_\text{eff}$ that depend on the spherical harmonic degree $\ell$ and the molecular diffusivities $\nu,\kappa$ according to \citep{Nataf2015}
\begin{eqnarray}
(\nu_\text{eff},\kappa_\text{eff})&=&(\nu,\kappa)~\mathrm{for}~\ell < \ell_{h}, \\
(\nu_\text{eff},\kappa_\text{eff})&=&(\nu,\kappa)\, q_{h}^{\ell-\ell_{h}} ~\mathrm{for}~\ell\ge \ell_{h}.
\end{eqnarray}
Here $\ell_{h}$ is the cut-off degree below which the hyperdiffusive treatment is not applied, and $q_{h}$ is the strength of hyperdiffusion. In Table \ref{bigtable1} we recall the values of $q_{h},\ell_{h}$ of a few LES cases from A17 \rev{(marked with a star symbol)}, where the approach was to start hyperdiffusivity at a relatively large scale $\ell_{h}=30$ and smoothly ramp its strength up towards smaller scales. Here we increase the model resolution $(\ell_{\mathrm{max}},NR)$ and test three approaches to relax hyperdiffusivity. First, we introduce new LES simulations where we keep $\ell_{h}=30$ but use weaker hyperdiffusivity, i.e. values of $q_{h}$ weaker than in A17. Second, we introduce quasi-DNS simulations (from hereafter qDNS) where we bring $\ell_{h}$ close to the smallest numerically resolved scale $\ell_{\mathrm{max}}$ and employ larger values of $q_{h}$ than those used in A17. This latter configuration attempts to maximise the range of length scales over which hyperdiffusivity is not applied, and capture as much as possible the physically relevant length scales within this range. Third, we extend the portion of the path where full DNS computations are carried out to 21 percent, by reporting on a DNS at $E=\te{-6}$ ($\epsilon=3.33~\te{-2}$). Due to the increased computational cost, qDNS and DNS simulations have been integrated over a shorter simulation time than their LES counterparts (see run length data in Table \ref{bigtable1}, presented in units of the overturn time $D/U$).

\begin{table}
\begin{center}
\setstretch{0.8}\small
\begin{tabular}{lrm{1.5cm}m{1.7cm}rrrrrr}
\hline\\[-0.5cm]
Case & $\epsilon$ & Path position (percent) & run length (overturns) & $E$ & $Pm$ & $\ell_\mathrm{max}$ & NR & $\ell_{h}$ & $q_{h}$\\
\hline
DNS (*)  &	0.1	& 14	 & 164 &	$3~\te{-6}$ & 0.8 & 256 & 480 & - & -\\			
DNS  &  $3.33~\te{-2}$ & 21 & 59 & $\te{-6}$ & 0.45 & 320 & 720 & - & -\\[0.3cm]			

qDNS & $3.33~\te{-2}$ & 21 & 8 & $\te{-6}$ & 0.45 & 320 & 720 & 290 & 1.1\\	
qDNS1 & $\te{-2}$ & 29 & 33 & $3~\te{-7}$ & 0.25 & 256 & 816 & 196 & 1.15\\	
qDNS2 & $\te{-2}$ & 29 & 3.3 & $3~\te{-7}$ & 0.25 & 512 & 1248 & 448 & 1.15\\	
qDNS & $3.33~\te{-4}$ & 50 &	0.8 & $\te{-8}$ & 0.045 & 640 & 2496 & 512 & 1.45\\[0.3cm]	

LES (*) & 0.1 & 14 & 1172 & $3~\te{-6}$ & 0.8 & 133 & 200 & 30 & 1.045\\	
LES (*) & $3.33~\te{-2}$ & 21 & 595 & $\te{-6}$ & 0.45 & 133	 & 240 & 30 & 1.0575\\	
LES1 (*) & $\te{-2}$ & 29 & 332 & $3~\te{-7}$ & 0.25 & 133 & 320 & 30 & 1.07 \\	
LES2 & $\te{-2}$ & 29 & 22.3 & $3~\te{-7}$ & 0.25 & 256 & 320 & 30 & 1.03\\		
LES (*) & $3.33~\te{-4}$ & 50 & 196 & $\te{-8}$ & 0.045 & 133 & 624 & 30 & 1.10\\	
\hline
\end{tabular}
\end{center}
\vspace*{0.1cm}\setstretch{1}\normalsize
\caption{\label{bigtable1} Input parameters of numerical models (see text for definitions). \rev{Cases marked with a (*) symbol have been previously reported in A17.}}
\end{table}

\begin{figure}
\centerline{\includegraphics[height=7cm]{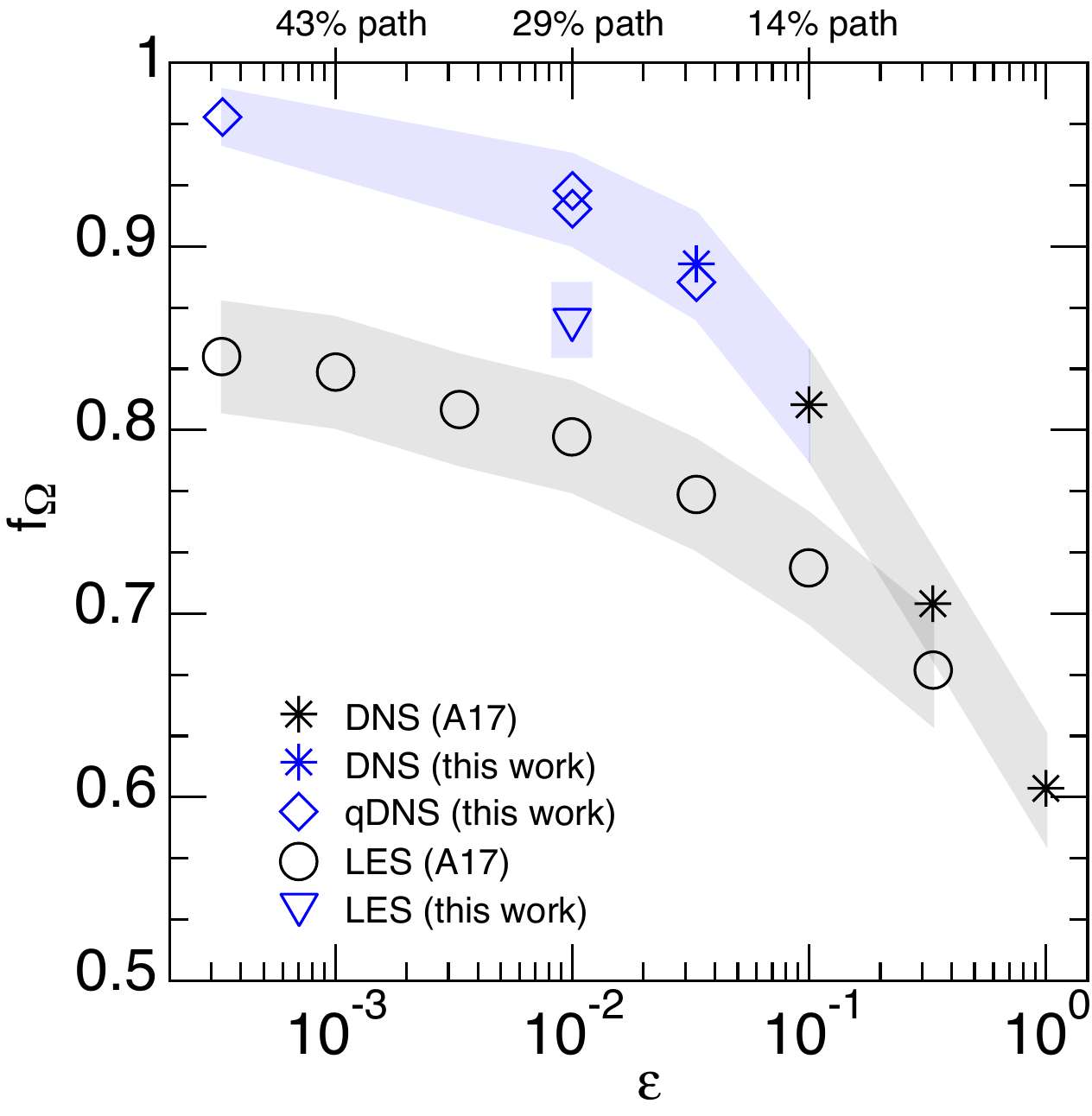}}
\caption{\label{fohm}Evolution of the fraction $f_\Omega$ of convective power dissipated by Ohmic losses with the path parameter $\epsilon$, in LES, qDNS and DNS simulations (Earth's core conditions are towards the left of this graph). Simulation results already reported in A17 are presented in grey, and new results are reported in blue. The shaded regions represent the $\pm 1$ std. dev. of fluctuations relative to the time average.}
\end{figure}

Time-averaged diagnostic outputs of the simulation are presented in Table \ref{bigtable2}. Listed first are the convective power density $p$ in units of $g_{o}F/4\pi D^{2}$, its fraction $f_\Omega$ that is dissipated by ohmic losses (also represented in Fig. \ref{fohm} as a function of the path parameter), the magnetic Reynolds number $Rm$ presenting the shell r.m.s. velocity $U$ in units of $\eta/D$, the shell r.m.s magnetic field $B$ in Elsasser units of $\sqrt{\rho\mu\eta\Omega}$ (where $\mu$ is the magnetic permeability, the resulting dimensionless value representing the square root of the classical Elsasser number, A17). We also report on the level of enforcement of the Taylor constraint by computing the level of cancellation $\mathcal{T}$ of azimuthal magnetic force integrated over geostrophic cylinders as in A17. The magnetic dissipation length scale $d_\Omega$ \rev{is defined} as in A17 (where it was labelled as $d_\mathrm{min}$)  \rev{from the square root of the ratio of volume-averaged magnetic energy and magnetic dissipation, and reported as an} equivalent harmonic degree $\ell_{\Omega}=\pi D/d_{\Omega}$. 

\begin{table}
\begin{center}
\setstretch{0.8}\small
\begin{tabular}{lm{1.5cm}rrrrrrrrrr}
\hline\\[-0.5cm]
Case & Path position (percent) & $\dfrac{4\pi D^{2} p}{g_{o} F}$ & $f_\Omega$ & $Rm=\dfrac{UD}{\eta}$ & $\dfrac{B}{\sqrt{\rho\mu\eta\Omega}}$ & $\mathcal{T}$ & $\ell_{\Omega}$ & $\ell_\mathrm{MS}$ & $\ell_{\perp}$  & $\ell_\mathrm{CIA}$ & $\ell_\mathrm{VAC}$\\ 
\hline
DNS (*) &  14 & 0.332 & 0.81 & 1092 & 4.52 &  0.17 &  159 & 71 &  10 & 76 & 207 \\
DNS  &   21 & 0.335 & 0.89 & 1071 & 4.41	&  0.11 & 166 & 70 & 10  & 110 & $>$320 \\[0.3cm]

qDNS &  21 &  0.331 & 0.88 & 1050 &	4.33 &  0.11 & 167 & 78 & 11 & 105 & 295\\
qDNS1 & 29  & 0.346 & 0.92 & 1130 &	4.49 &  0.07 & 171 & 60 & 10 & 199 & 212 \\
qDNS2 &  29 & 0.340 & 0.93 & 1095 &	4.49 &  0.07 & 170 & 65 & 12 &  182 & 455 \\
qDNS &  50 & 0.343 & 0.97 &  1105   &        3.90       & 0.02  & 196 & 130 & 13 & 522 & 522 \\[0.3cm]

LES (*) & 14 & 0.336 & 0.72 &	1046 & 4.64 & 0.17  & 146 & 62& 9 & 61 & 73 \\
LES (*) &  21 & 0.343 & 0.76 & 1036	 & 4.57 & 0.12 & 150 & 65 & 10 & 72 & 79 \\
LES1 (*) & 29 & 0.348 & 0.80 &	1046 & 4.70 &  0.08 & 152 & 60 & 10 & 87 & 83 \\
LES2 & 29 	& 0.347 & 0.86 &	 1084 & 4.61 &  0.09 & 162 & 67 & 12 & 122 & 127 \\
LES (*) & 50 	& 0.360 & 0.84 &	 1089 & 4.48 &  0.04 & 164 & 81 & 12  & 101 & 92 \\
\hline
\end{tabular}
\end{center}
\vspace*{0.1cm}\setstretch{1}\normalsize
\caption{\label{bigtable2} Output parameters of numerical models (see text for definitions). \rev{Cases marked with a (*) symbol have been previously reported in A17.}}
\end{table}

\begin{figure}
\centerline{\includegraphics[width=17cm]{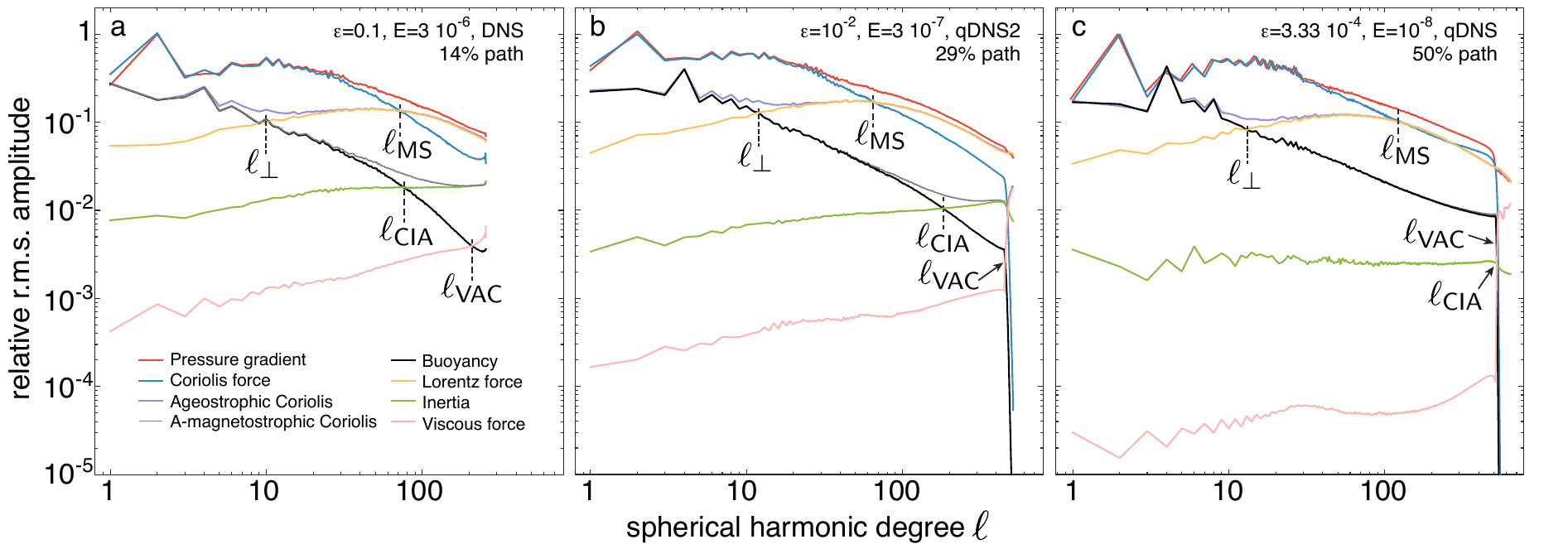}}
\caption{\label{DNSforcebal} Scale-dependent force balance diagrams for direct and quasi-direct numerical simulations (DNS and qDNS) obtained along the parameter space path. Represented are the r.m.s. amplitudes of each force as functions of the spherical harmonic degree $\ell$ (see A17 for technical computation details). Forces are normalised relative to the maximum value of the pressure force. The ageostrophic Coriolis force represents the residual Coriolis force after removal of the pressure force. The amagnetostrophic Coriolis force represents the residual Coriolis force after removal of the pressure and Lorentz forces. The crossover length scales $\ell_\mathrm{MS}$, $\ell_{\perp}$, $\ell_\mathrm{CIA}$ and $\ell_\mathrm{VAC}$ are defined from the crossings observed in these diagrams (see text for interpretation, values in Table \ref{bigtable2} and a summary diagram in Fig. \ref{scales}).}
\end{figure}

\begin{figure}
\centerline{\includegraphics[width=12cm]{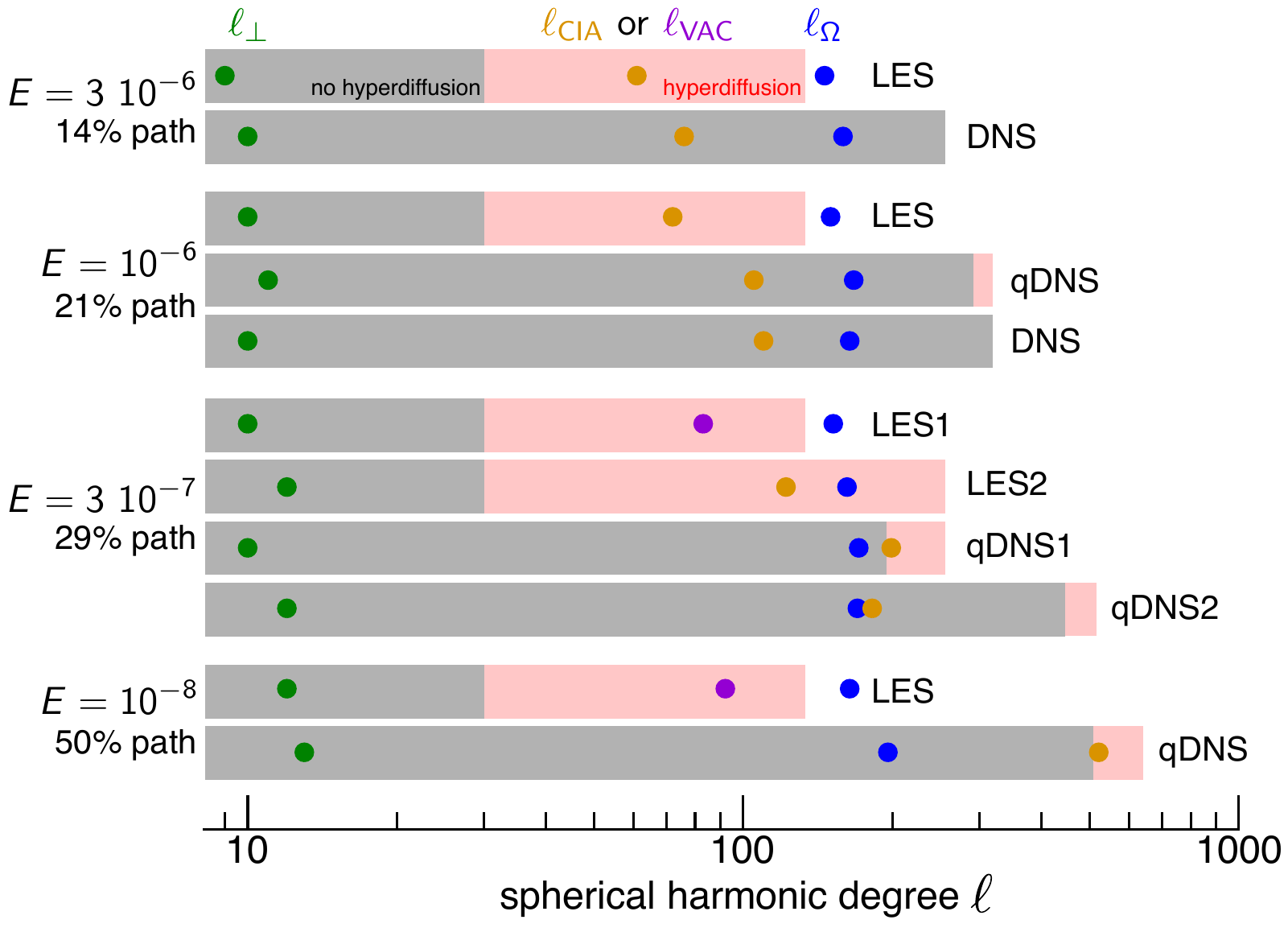}}
\caption{\label{scales}Relative positions of key length scales in the simulations, represented as equivalent spherical harmonic degrees. The horizontal bands represent the extent of the spherical harmonics expansion (with the right end representing the maximum degree $\ell_{\mathrm{max}}$), and their color delineate the ranges where hyperdiffusivity is applied (pink) or not (grey). Note that $\ell_{\Omega}$ is based on the integral length scale $d_{\Omega}$ encompassing magnetic diffusion in all three spatial dimensions, hence it is possible that $\ell_{\Omega}$ exceeds $\ell_{\mathrm{max}}$. The degree $\ell_{\perp}$ is indicative of the large scale at which the first-order MAC balance is achieved. Only the smallest of the two harmonic degrees $\ell_\mathrm{CIA}$, $\ell_\mathrm{VAC}$ is represented, thereby indicating the nature of the force balance achieved at second order.}
\end{figure}

As previously in A17, we compute scale-dependent force balance diagrams (Fig. \ref{DNSforcebal}) that enable the investigation of the spatial force balance structure. To represent length scales, we prefer the harmonic degree $\ell$ over the harmonic order because of the  invariance of $\ell$-dependent energy spectra against rotation of the coordinate system\rev{, and because unlike an harmonic order, a given degree $\ell$ in the spectral space can be associated to a minimal corresponding length scale $\pi D/\ell$ in the physical space. It has been common practice to use curled forces rather than the forces themselves to examine their contributions in the Navier-Stokes equation \citep[e.g.][]{ChristensenAubert2006,Davidson2013}. This approach is however misleading if these curls are then analysed in the spectral space. Indeed, the contributions at different harmonic degrees are coupled by the curl operation, thereby defeating the possibility to associate a force at degree $\ell$ to its action at length scale $\pi D/\ell$ in the physical space. For this reason, and also because we wish to also assess the role of the pressure force, it is mandatory to analyse the non-curled forces in the spectral space (as previously done in A17).} Crossover length scales are obtained in Fig. \ref{DNSforcebal} and reported in Table \ref{bigtable2} by determining the harmonic degrees at which two forces of the system become of equal amplitude. These may be thought of as scales where a given force balance is achieved, and they usually bound scale ranges where different dynamical equilibria are enforced. The temporal variability of force balance diagrams is weak (A17) and they are therefore computed from snapshots in time. The typical error on crossover length scales is then less than 10 percent. Here we recall some key results found in A17 for the force balance structure and introduce new crossover length scales. \rev{These results found along the parameter space path are typical of the situation most commonly found in a wide region of the parameter space \citep{Schwaiger2019}, with the exceptions being found at low convective supercriticality and low magnetic Prandtl number $Pm$.} The different forces span up to five orders in magnitude (with unprecedently low levels of viscosity being reached in the qDNS at 50 percent of the path, Fig. \ref{DNSforcebal}c), involving at least three successive orders of force balances. Fig. \ref{DNSforcebal} shows that at large scales (low harmonic degrees), a leading order quasi-geostrophic balance is achieved between the Coriolis and pressure forces. The crossover of the Coriolis and Lorentz forces defines the magnetostrophic harmonic degree $\ell_\mathrm{MS}$. For harmonic degrees larger than $\ell_\mathrm{MS}$, the zeroth-order balance becomes magnetostrophic, with the pressure force equilibrating the Lorentz force (which is hence essentially a magnetic pressure) and the residual Coriolis force. Considering now the first-order balance, the crossover harmonic degree of the Lorentz and buoyancy forces is defined as $\ell_{\perp}$. As this balance is achieved under the constraint of the part of the Coriolis force not balanced at zeroth order by pressure (the ageostrophic Coriolis force), the associated length scale is usually referred to as that of the MAC (magneto-Archimedes-Coriolis) balance \citep[][A17]{Davidson2013,Yadav2016PNAS,Aubert2018}. For $\ell<\ell_{\perp}$, the first order balance is of thermal wind type between the ageostrophic Coriolis and buoyancy forces, and for $\ell>\ell_{\perp}$ it becomes of magnetic wind type between the ageostrophic Coriolis and Lorentz force. Moving to the next-order force balances, the crossover harmonic degrees of the inertial and buoyancy forces $\ell_\mathrm{CIA}$, and of the viscous and buoyancy forces $\ell_\mathrm{VAC}$ are respectively representative of the CIA (Coriolis-inertia-Archimedes) and VAC (viscous-Archimedes-Coriolis) force balances that operate under the constraint of the Coriolis force residual not balanced at zeroth and first orders (the amagnetostrophic Coriolis force). A summary diagram presenting the relative positions of the main length scales is presented on Fig. \ref{scales}. Scale separation in this system refers to the difference between the large scale of the first-order MAC balance at degree $\ell_{\perp}$, and the much smaller scales of the second-order force balances at degrees $\ell_\mathrm{CIA}$, $\ell_\mathrm{VAC}$ and of magnetic dissipation at degree $\ell_{\Omega}$.

\subsection{\label{QGMAC}Asymptotic QG-MAC theory}
The results from the upsized numerical simulations will be analysed in the light of the asymptotic theory proposed by \cite{Davidson2013}, some elements of which are recalled here. We consider a leading-order, quasi-geostrophic (QG) equilibrium and a first-order magneto-Archimedes-Coriolis (MAC) force balance, both well supported by the numerical simulations (Fig. \ref{DNSforcebal}), together with the dynamo power budget. Due to the zeroth-order QG equilibrium and the associated Taylor-Proudman constraint, the scale $d_{\parallelsum}$ of flow structures along the rotation axis $\vc{\Omega}$ remains at the system size i. e. $d_{\parallelsum}\sim D$ (see Fig. \ref{Vviz}). The first-order MAC balance operating at the large scale $d_{\perp}=\pi D/\ell_\perp$ can be written from the curl of the Navier-Stokes equation (see A17):
\begin{equation}
\dfrac{\rho\Omega U}{D} \sim \dfrac{g_{o} C}{d_{\perp}} \sim \dfrac{B^{2}}{\mu d_{\perp}^{2}},\label{MACbal}
\end{equation} 
where $C$ is a typical value for the density anomaly field. In the present context where Ohmic losses account for the essential part of the dissipated energy (Fig. \ref{fohm}), the balance between the convective power input and magnetic dissipation \citep{Christensen2004,ChristensenAubert2006} may be written
\begin{equation}
\dfrac{\eta B^{2}}{\mu d_\Omega^2} \sim p\sim \dfrac{g_{o} F}{D^{2}},\label{Ebudget}
\end{equation}
where we have used the equivalence between convective power and mass anomaly flux \citep[][A17]{ChristensenAubert2006,Aubert2009}, as attested by the \rev{almost} constant value of $4\pi D^{2} p/ g_{o} F$ in Table \ref{bigtable2}. Combining equations (\ref{MACbal}) and (\ref{Ebudget}), and using $F\sim U C D^2$, we obtain the vorticity equivalence
\begin{equation}
U/d_{\perp}= c_{\omega} ~\eta/d_\Omega^{2}.\label{vorteq}
\end{equation}
This equivalence states that the vorticities at the large scale $d_\perp$ where the magnetic field is sustained by drawing convective power (balance between buoyancy and Lorentz forces) and at the small scale $d_\Omega$ where this power is ohmically dissipated are related by a proportionality constant $c_{\omega}$ \rev{(interpreting $\eta/d_\Omega^{2}$ as a vorticity implicitly assumes that the gradient and rotational parts of $\vecu$ are of similar magnitude at the scale of magnetic diffusion).}

\subsubsection{\label{powerlaw}Self-similar ranges in spatial spectra}
Despite the fact that power is essentially carried by magnetic (rather than hydrodynamic) non-linearities from the large injection scale $d_\perp$ down to the small dissipation scale $d_\Omega$, we do not expect spatial spectra of the magnetic energy density to feature 
self-similar, power-law behaviours in the range $[\ell_\perp,\ell_\Omega]$ because the level of magnetic turbulence is only moderate ($Rm$ is moderate, Table \ref{bigtable2}). However, the hydrodynamic Reynolds number $Re=Rm/Pm$ reaches values up to 24000 along the explored part of the path, suggesting that such ranges are likely in spectra of the velocity field. The vorticity equivalence  (\ref{vorteq}) further supports this idea. Indeed, the constant $c_\omega$ being in principle of order 1, at large values $d_\perp/d_\Omega$ of the scale separation we expect the spectral enstrophy density $\omega^{2}(\ell)$ at degree $\ell$ (where $\vecom=\nabla\times\vecu$ is vorticity) to remain approximately constant i.e. 
\begin{equation}
\omega^{2}(\ell) \propto \ell^{0}~\mathrm{for}~\ell_{\perp}<\ell<\ell_{\Omega}.\label{omegalaw}
\end{equation}
With length scales varying like $\ell^{-1}$, the energy density of velocity $u^{2}(\ell)$ should then also present a power-law range
\begin{equation}
u^{2}(\ell) \propto \ell^{-2}~\mathrm{for}~\ell_{\perp}<\ell<\ell_{\Omega}.\label{ulaw}
\end{equation}
Deviations from these ideal spectra are however likely. First, length scales are indeed anisotropic in this system, with flow structures elongated along the rotation axis ($d_{\perp} \ll d_{\parallelsum}$), implying that the velocity energy spectrum is less steep than $\ell^{-2}$ if the vorticity spectrum is flat. In that sense, the spectra of the vorticity component parallel to the rotation axis $\omega_{\parallelsum}=\left(\nabla\times\vecu\right)\cdot \vc{\Omega}/\Omega$ and velocity component perpendicular to the rotation axis $\vecu_{\perp}=\vecu-\left(\vecu\cdot\vc{\Omega}/\Omega\right)$ should better approach (\ref{omegalaw},\ref{ulaw}) than those of the full vorticity $\vc{\omega}$ and velocity $\vecu$, because the former leave out the variations along the rotation axis. Second, the scale separation $d_\perp/d_\Omega$ only reaches values up to 20 in our simulations (Fig. \ref{scales}). Finally, the correspondence between length scales and harmonic degrees is not straightforward \citep[e.g.][]{Schaeffer2017}, because a given harmonic degree can represent different length scales at different radii, and because different dynamics may exist in the regions inside and outside the axial cylinder tangent to the inner core (the tangent cylinder). We will nevertheless use (\ref{omegalaw},\ref{ulaw}) as guidelines for the interpretation of spatial spectra.

\subsubsection{\label{asympt}Asymptotic scaling laws along the parameter space path}
We now recall the derivation of the scaling laws for $U$, $B$, $d_{\perp}$, and $d_{\Omega}$ along the parameter space path, as functions of the path parameter $\epsilon$. If one requires the large- and small-scale vorticities present in equation (\ref{vorteq}) to be independent on the system rotation rate and diffusivities, then from (\ref{Ebudget}) it follows that the magnetic field itself is independent on the rotation rate and diffusivities. \cite{Davidson2013} pointed out that dimensional analysis finally yields the following scaling, corresponding to the initial proposal of \cite{ChristensenAubert2006}:
\begin{equation}
B \sim \sqrt{\rho\mu} (g_{o}F/\rho D)^{1/3}.\label{CAscaling}
\end{equation}
Combining equations (\ref{MACbal},\ref{Ebudget},\ref{CAscaling}) and using $Ra_{F}\sim\epsilon$ leads to the scalings previously proposed in A17, which are presented here in a slightly different manner to account for our choice of dimensionless quantities:
\begin{eqnarray}
\dfrac{B}{\sqrt{\rho\mu\eta\Omega}} \sim \epsilon^{1/12},\label{Bscal}\\
Rm=\dfrac{UD}{\eta}\sim \epsilon^{-1/18},\label{Uscal}\\
d_{\perp}/D \sim \epsilon^{1/9},\label{dperpscal}\\\
d_{\Omega}/D \sim \epsilon^{1/12}.\label{dohmscal}
\end{eqnarray}
The powers of $\epsilon$ entering (\ref{Bscal}-\ref{dohmscal}) imply that these quantities only undergo weak variations along the path (as confirmed by Table \ref{bigtable2}), from $\epsilon=1$ down to $\epsilon=\te{-7}$. This is another evidence of the continuum that exists along the parameter space path between numerical simulations and Earth's core conditions. In particular, the weak variations of the length scales $d_{\perp}$ and $d_{\Omega}$ (see Fig. \ref{scales}) led A17 to propose a scale-invariant theory \rev{\citep[the path theory, see also][]{Starchenko2002}} tailored for the analysis of LES simulations, where these two length scales are fixed, which in turn implies constant values of $Rm$ and $B/\sqrt{\rho\mu\eta\Omega}$. The weak values of scaling exponents may initially obscure the prospect of checking these theories against numerical data, but A17 have shown that this is not the case owing to the broad range of accessible path parameter values $\epsilon$ and the low scatter of numerical data. However, an uncertainty analysis will be needed for the upsized runs, which have a short integration time (Table \ref{bigtable1}) and may therefore present more scatter than the LES simulations analysed in A17.

\section{\label{results}Results}

\subsection{Transients}
\begin{figure}
\centerline{\includegraphics[width=8cm]{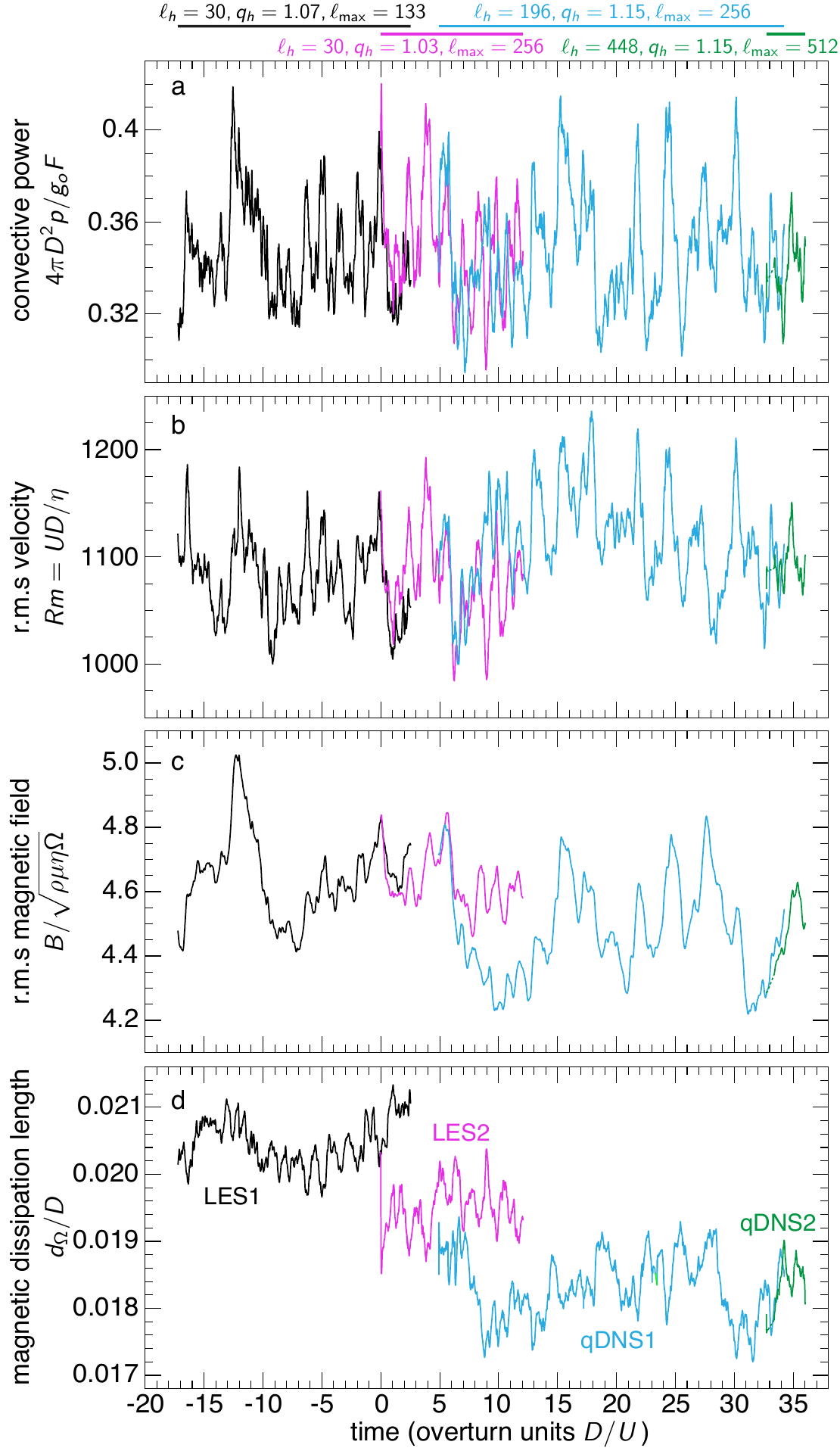}}
\caption{\label{transients}Temporal evolution of output diagnostics during an upsizing sequence at 29 percent of the parameter space path ($E=3~\te{-7}$). At each upsizing step, the restart file from the previous step is used. Case LES1 from A17 (Table \ref{bigtable1}, black)  is first expanded into LES2 (purple) with the same cut-off harmonic degree $\ell_{h}=30$ but a lower value of the hyperdiffusion strength $q_{h}$. Two quasi-DNS simulations (qDNS1, qDNS2) are then performed, first by expanding the spherical harmonic truncation to $\ell_\mathrm{max}=256$ and the hyperdiffusivity cut-off to $\ell_{h}=196$ (blue), and then by expanding $\ell_\mathrm{max}$ to 512 and $\ell_{h}$ to 448. For clarity, only a portion of the first three sequences (black, purple, blue) is represented. \rev{Because of an incorrect specification of the numerical code options, the diagnostic outputs presented here were} not available during a short portion of the last sequence (green dashes).}
\end{figure}

Fig. \ref{transients} presents a typical temporal evolution sequence of output diagnostic quantities during a gradual upsizing process starting from LES1 (Table \ref{bigtable1}), a large-eddy simulation at 29 percent of the parameter space path (Ekman number $E=3~\te{-7}$) taken from A17. At each increase of spatial resolution, the simulation is restarted using the output of the previous step. 
Transients associated to this process are short, typically a fraction of a convective overturn. As we shall see, the process does not alter the leading-order QG-MAC force balance (Fig. \ref{DNSforcebal}, see also Fig. \ref{tql} below) and large-scale simulation structure (see Figs. \ref{Vviz}-\ref{Bviz}), leaving only the smaller scales to reach their saturated energy after the upsizing step. Integral diagnostics for power, velocity and magnetic field (Fig. \ref{transients}a-c) are therefore largely invariant against upsizing, with the upsized simulations even following the evolution of the previous lower-resolution case for about an overturn time after the upsizing step. The magnetic dissipation length scale $d_{\Omega}$ (Fig. \ref{transients}d) is more sensitive to the distribution of small scales and hence more influenced by upsizing, until a high enough resolution is reached in qDNS2 to fully encompass $\ell_{\Omega}=\pi/d_{\Omega}$ (see Fig. \ref{scales}). \rev{As a corollary, the fraction $f_{\Omega}$ of Ohmic dissipation in the system (Fig. \ref{fohm}) generally increases during the upsizing procedure, and reaches a value independent on further upsizing when $d_{\Omega}$ also converges (see similarity of $f_{\Omega}$, $\ell_{\Omega}$ for qDNS1 and qDNS2 in Table \ref{bigtable2}). At 50 percent of the path, the upsizing process thereby leads to a value  $f_{\Omega}=0.97$ indicative of almost entirely Ohmic dissipation.} We are therefore confident that despite being integrated for a small amount of overturn times, the upsized simulations are fully representative of the saturated dynamical state that could be reached at much greater computational cost in a regular DNS. 

\subsection{Planforms in physical space}
\begin{figure}
\centerline{\includegraphics[width=14cm]{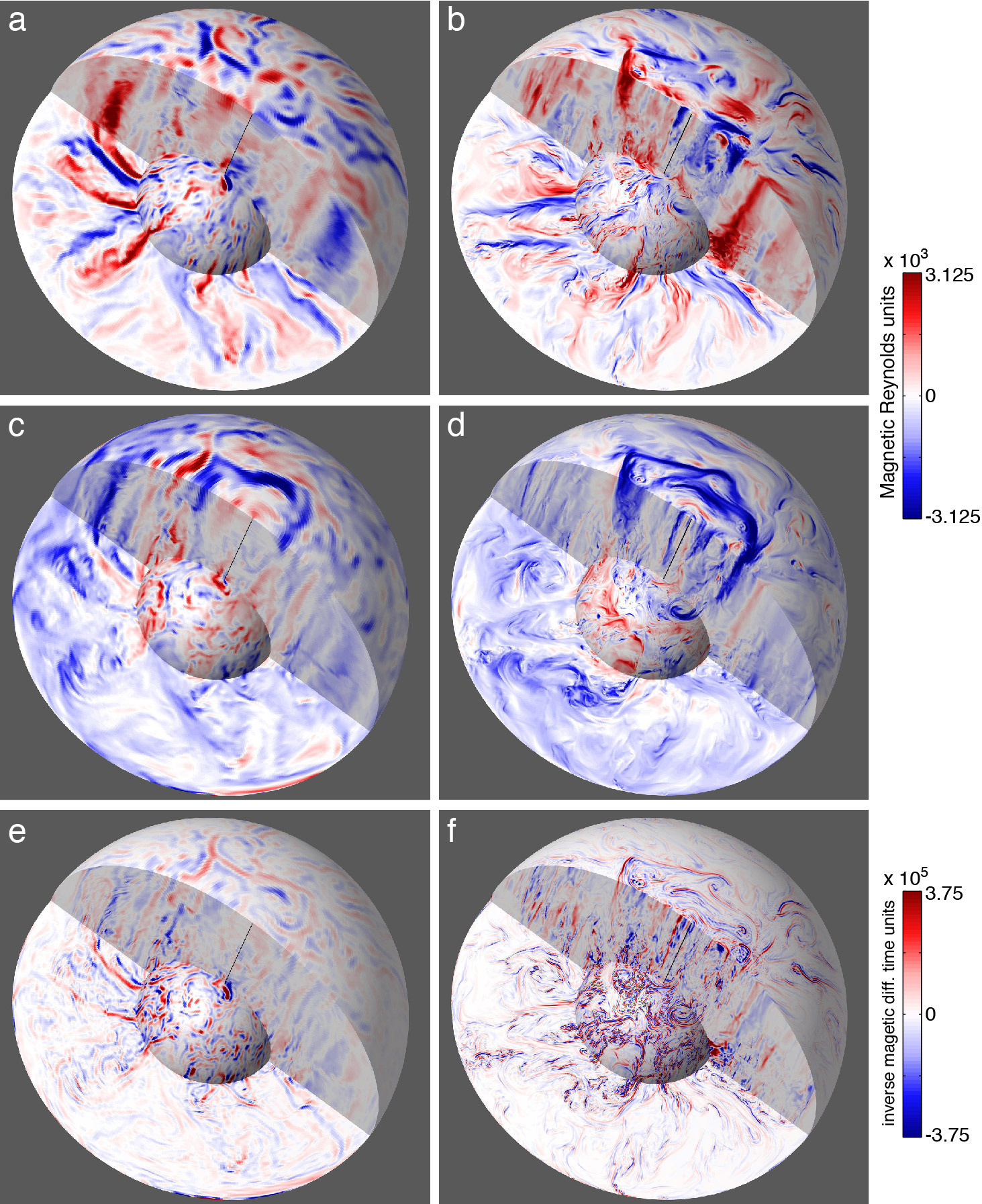}}
\caption{\label{Vviz} Planforms of velocity $\vecu$ components in the cylindrical radial direction (a,b), the azimuthal direction (c,d), and planforms of the axial vorticity $\omega_{\parallelsum}$ (e,f). Represented are the fields in the equatorial plane, in a half meridian plane, at the outer (core-mantle) boundary of the shell at radius $r_{o}$, and at radius $r/D=0.545$, close to the inner (inner core) boundary. The position of the rotation axis $\vc{\Omega}$ is marked by a black line on all panels. The left column (a,c,e) refers to a snapshot taken during case LES1 from A17 at 29 percent of the path (Ekman number $E=3~\te{-7}$, Table \ref{bigtable1}), and the right column to a snapshot of the highest-resolution case qDNS2 also obtained at 29 percent of the path after the upsizing sequence (green lines in Fig. \ref{transients}). Velocity is presented in magnetic Reynolds units of $\eta/D$, and vorticity in inverse magnetic diffusion time units of $\eta/D^{2}$.}
\end{figure}

\begin{figure}
\centerline{\includegraphics[width=14cm]{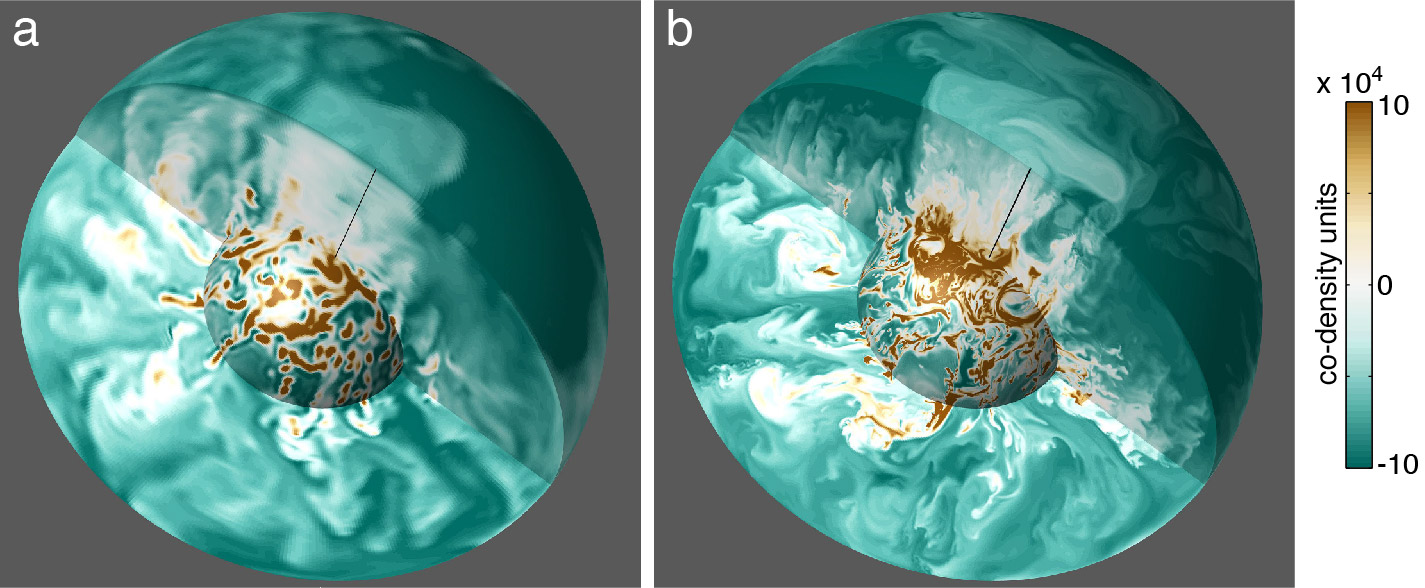}}
\caption{\label{Cviz}Planforms of the density anomaly field $C$, presented in units of $F/4\pi\Omega D^3$. The snapshots are taken at the same instants in time as in Fig. \ref{Vviz}, for cases LES1 (a) and qDNS2 (b) obtained at 29 percent of the path. Same visualisation conventions as in Fig. \ref{Vviz}.}
\end{figure}

\begin{figure}
\centerline{\includegraphics[width=14cm]{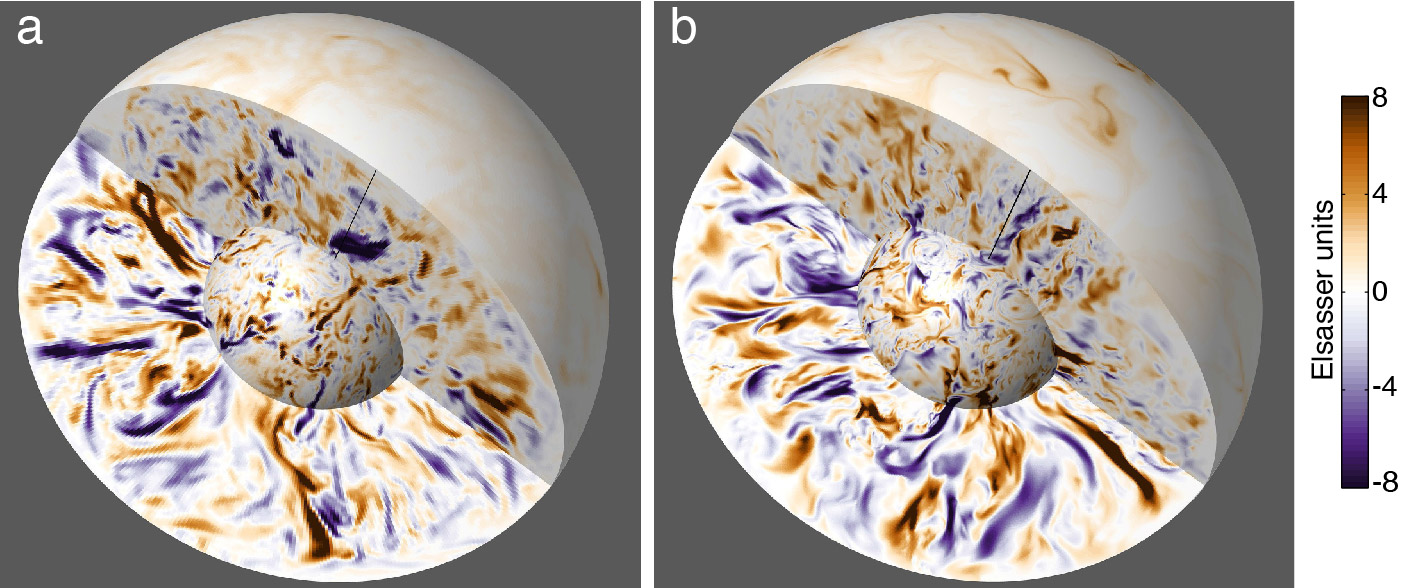}}
\caption{\label{Bviz}Planforms of the radial component of the magnetic field $\vc{B}$, presented in Elsasser units of $\sqrt{\rho\mu\eta\Omega}$. The snapshots are taken at the same instants in time as in Fig. \ref{Vviz}, for cases LES1 (a) and qDNS2 (b) obtained at 29 precent of the path. Same visualisation conventions as in Fig. \ref{Vviz}.}
\end{figure}

\begin{figure}
\centerline{\includegraphics[width=12cm]{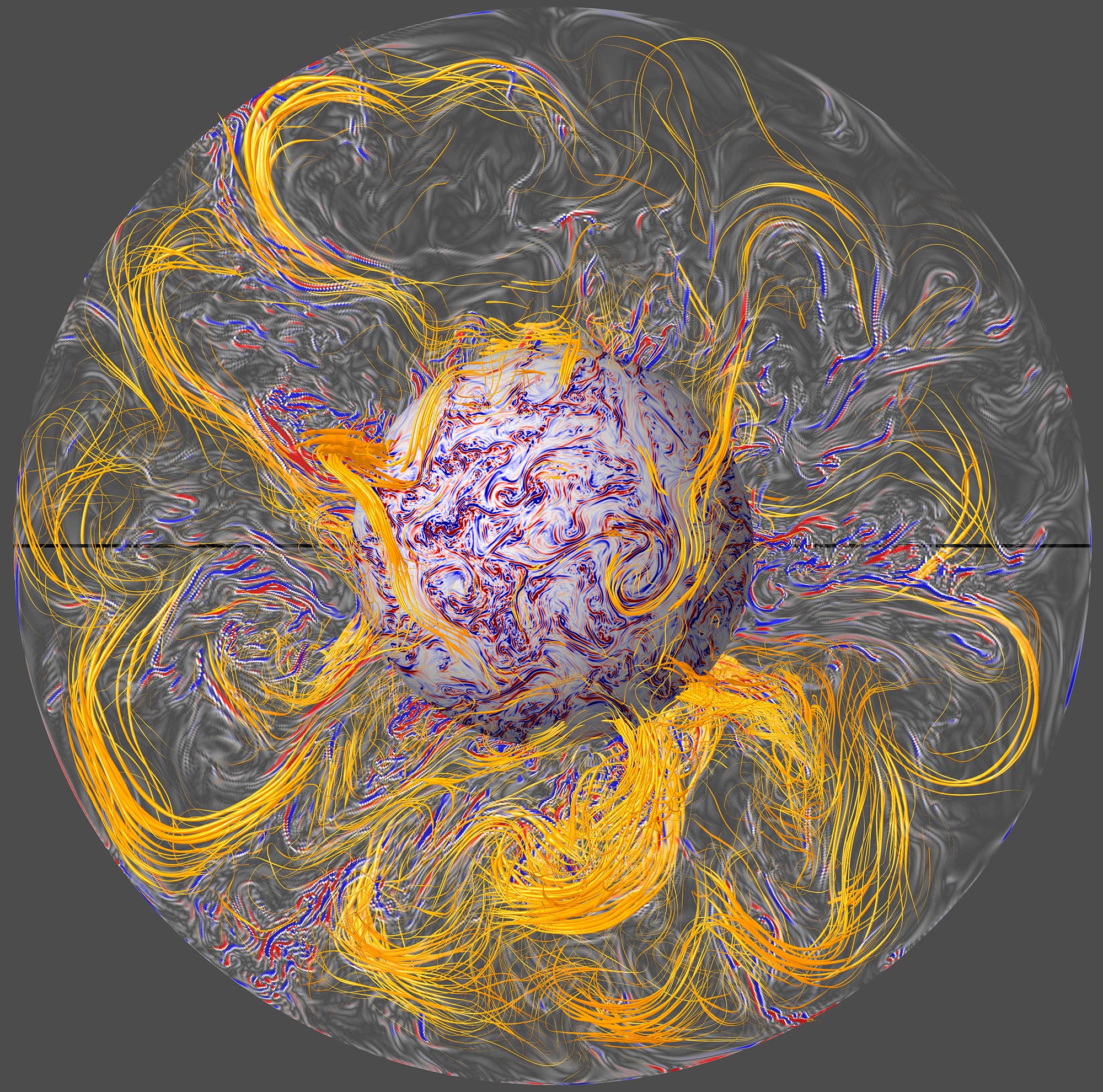}}
\caption{\label{align} Polar view of the axial vorticity field $\omega_{\parallelsum}$ from case qDNS2 at 29 percent of the path, presented in the entire equatorial plane and at radius $r/D=0.545$, with black lines delineating the cut presented in Fig. \ref{Vviz}f (same snapshot and color coding as this figure). Magnetic field lines parallel to the local direction of $\vc{B}$ are also represented (orange), with thickness proportional to the local magnetic energy density $\vecB^{2}$. In order to see field lines in the whole volume, the equatorial plane is made selectively transparent with opacity proportional to $|\omega_{\parallelsum}|$. This reveals an organisation of field lines aligned with the columnar vorticity filaments and along the rotation axis.}
\end{figure}

Planforms of the velocity, density anomaly and magnetic fields in the physical space are presented in Figs. \ref{Vviz}-\ref{Bviz}, for the initial case LES1 and final case qDNS2 of the upsizing sequence at 29 percent of the path (Table \ref{bigtable1}, Fig. \ref{transients}). The structure of $\vecu,\vecB,C$ at large scales is remarkably preserved after upsizing, confirming the accuracy of LES, as previously advocated for in A17. These large scales consist in upwellings taking the shape of radially elongated columnar sheets (Fig. \ref{Vviz}a,b) associated with large buoyancy plumes (Fig. \ref{Cviz}a,b), entrained in a mostly columnar and westward azimuthal flow presenting rim-like polar circulations on the tangent cylinder (Fig. \ref{Vviz}c,d). Relative to its LES1 counterpart, the qDNS2 simulation however features significantly enriched small scales and a higher level of separation between the largest and smallest observable scales. These small scales comprise columnar flow filaments coexisting with the sheet-like cylindrical radial upwellings (Fig. \ref{Vviz}b) and ramifications of the buoyancy anomaly plumes (Fig. \ref{Cviz}b). The small-scale content is best seen on planforms of the axial vorticity $\omega_{\parallelsum}$ (Fig. \ref{Vviz}f) where the columnar filaments are gathered at the edges of the large-scale velocity structures. As we shall see below in Figs. \ref{spectratql},\ref{evolpath},\ref{UOm}, the large scale in $\vecu$, $\vecB$, $C$ is $d_{\perp}$, the scale of the MAC balance at which the magnetic field draws its power from convection, and the dominant small scale seen in $\vecom$ in the qDNS2 case corresponds to $d_{\Omega}$ where this power is ohmically dissipated. In the LES1 case (Figs. \ref{Vviz}a,c,d, \ref{Cviz}a), the scale separation is much weaker, and we shall see the small scale of $\vecom$ in this case correspond to the balance of hyperdiffusive viscosity and buoyancy force (the VAC balance at harmonic degree $\ell_\mathrm{VAC}$). Fig. \ref{Bviz}a,b shows that the upsizing process does not cause an increase in the scale separation of the magnetic field $\vecB$, the content of which remains largely at scale $d_{\perp}$. Indeed, in case LES1 the magnetic field is already close to being fully resolved, because the low value of the magnetic Prandtl number $Pm$ warrants a modest degree of magnetic turbulence $Rm$ at strong forcing. Analysing the magnetic field geometry of the upsized case qDNS2 (Fig. \ref{align}) it can be seen that field lines tend to avoid regions of strong velocity and also gather at the edges of upwelling sheets or density anomaly plumes. This leads to a remarkable alignment between small-scale vorticity filaments and magnetic field lines, which is commonly referred to as dynamic alignment in the theory of strong magnetohydrodynamic turbulence \citep[see][]{Tobias2012}. This is the process through which the magnetic field adjusts so as to minimise its shear by the flow, as a consequence of Lenz law. The Lorentz force therefore constrains the flow but ultimately becomes dynamically irrelevant, \rev{reducing} its contribution at small scales to a magnetic pressure (Fig. \ref{DNSforcebal}). The planforms of qDNS2 are broadly comparable to those of recently published DNSs with strong scale separation \citep{Schaeffer2017,Sheyko2018}. However, less accumulation of light material \rev{within the tangent cylinder} is observed in qDNS2 than in simulation S2 from \cite{Schaeffer2017}, and we do not observe the generation of reverse surface magnetic flux inside this cylinder \citep[compare Figs. \ref{Cviz}b, \ref{Bviz}b to Fig. 7 in][]{Schaeffer2017}. Case qDNS2 indeed features a stronger level of forcing (stronger hydrodynamic and magnetic Reynolds numbers) than the cases from \cite{Schaeffer2017,Sheyko2018} and this presumably improves the exchange of material through the tangent cylinder, leading to a spatially more homogeneous system. 

\subsection{Force balances at leading and first order}
\begin{figure}
\centerline{\includegraphics[width=17cm]{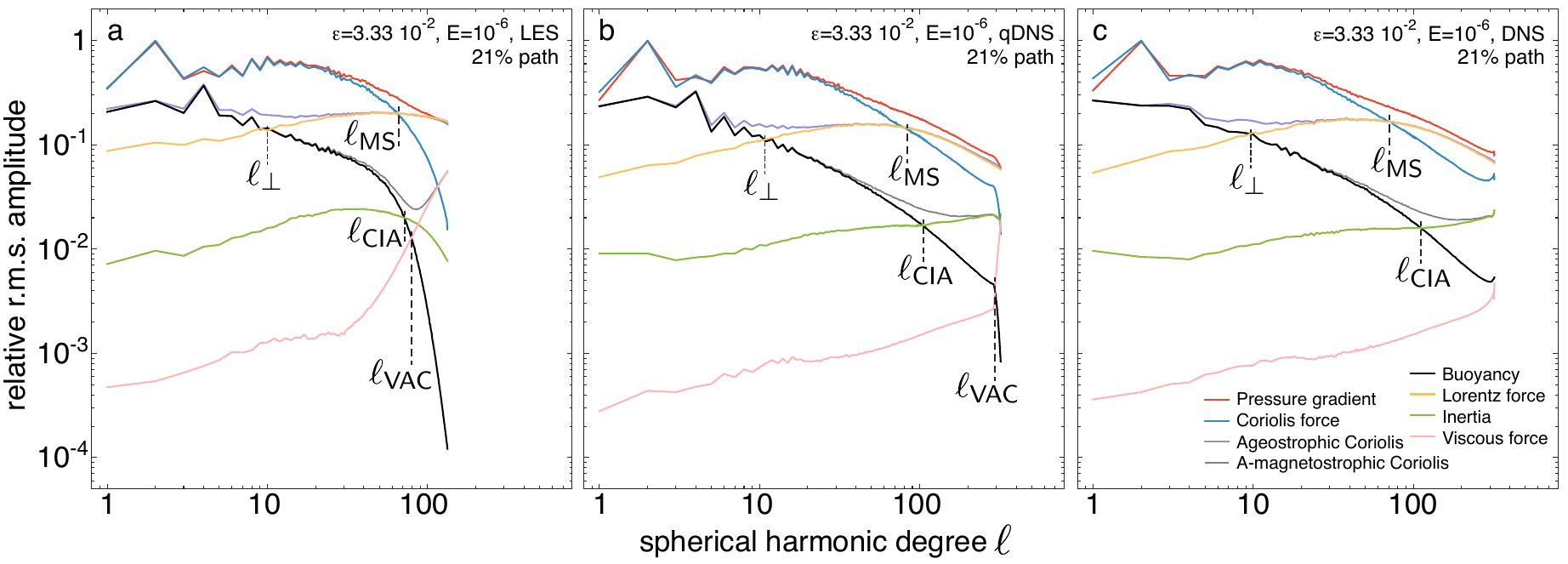}}
\caption{\label{tql}Scale-dependent force balances for LES (a), qDNS (b) and DNS (c) simulations at 21 percent of the parameter space path (Ekman number $E=\te{-6}$). See Fig. \ref{DNSforcebal} and text for the definitions of crossover harmonic degrees, and A17 for technical computation details.}
\end{figure}

Fig. \ref{tql} presents scale-dependent force balances in simulations with variable degrees of upsizing (LES, qDNS, DNS) at a fixed position on this path (21 percent). This complements Fig. \ref{DNSforcebal} where force balances are presented for DNS and qDNS along the parameter space path. Invariance of the zeroth-order quasi-geostrophic and magnetostrophic balances, as well as the first-order MAC balance was observed along the path in the LES of A17, and is confirmed here in the upsized qDNS and DNS. The LES are furthermore efficient at accurately capturing these balances (compare Fig. \ref{tql}a and \ref{tql}c). The value of the crossover length scale to magnetostrophy $\ell_\mathrm{MS}$ is also broadly invariant in our sets of runs, except for the qDNS at 50 percent of the parameter space path (Fig. \ref{DNSforcebal}c), where the force balance is perturbed by the choice of upsizing strategy (see below). Invariance of $\ell_\mathrm{MS}$ is consistent with the analysis of \cite{Aurnou2017} where this scale is conjectured to depend solely on the Elsasser number $B^{2}/\rho\mu\eta\Omega$ and magnetic Reynolds number $Rm$, both of which are about constant along the path (Table \ref{bigtable2}). The MAC length scale $\ell_{\perp}$ is also almost invariant along the parameter space and against upsizing (see also Fig. \ref{scales}). In summary, the QG-MAC force balance previously advocated for in \cite{Davidson2013,Yadav2016PNAS}; A17 ; \cite{Calkins2018} is therefore confirmed here as the leading-order balance asymptotically holding in rapidly rotating regimes, and it is shown to be remarkably stable even in the high-resolution, upsized simulations. 

The projection of these leading and first-order force balances on axisymmetric scales yields the Taylor constraint, stating that the integral of azimuthal Lorentz force over imaginary cylinders co-axial to the rotation axis vanishes to the level of the residual inertia and viscosity (see A17). Since the magnetic field structure is largely preserved by upsizing (Fig. \ref{Bviz}a,b) the level of enforcement $\mathcal{T}$ of the Taylor constraint is also preserved (Table \ref{bigtable2}) and increases along the parameter space path. We again emphasize that adherence to the Taylor constraint should not be misinterpreted as a sign of leading-order magnetostrophy, as it follows here from leading-order quasi-geostrophy.

\subsection{Second-order force balance and optimal approximations}

Differences between upsizing strategies emerge at the next orders in the force balance diagrams. DNS simulations (Figs. \ref{DNSforcebal}a, \ref{tql}c, see also A17) show that the second order force balance is of the CIA nature. It has been argued that the VAC balance could influence the dynamics of dynamo simulations at moderate parameter values \citep{King2013} but this is clearly not the case here as this balance only comes at the third order in amplitude. Quasi-DNS simulations constructed such that the hyperdiffusive cut-off $\ell_{h}$ largely exceeds $\ell_\mathrm{CIA}$ (such as qDNS at 21 percent and qDNS2 at 29 percent of the path, Fig. \ref{scales}) have a force balance structure that is undistinguishable from that of the full DNS down to the second order in amplitude (compare Fig. \ref{tql}b and \ref{tql}c, see also Fig. \ref{DNSforcebal}b). At third order, the VAC balance length scale $\ell_\mathrm{VAC}$ is improperly resolved in these qDNS but this has no consequence on the partitioning of dissipation between Ohmic and viscous losses (compare the $f_{\Omega}$ values in qDNS and DNS at 21 percent of the path in Fig. \ref{fohm}). This type of qDNS may therefore be safely used to fully replace a DNS, but it still involves a sizeable computational cost because $\ell_\mathrm{CIA}$ becomes large along the path (Fig. \ref{scales}). As an acceptable tradeoff between cost and accuracy, a qDNS simulation where the cut-off is close to, or smaller than the true value of $\ell_\mathrm{CIA}$, but still larger than $\ell_{\Omega}$ may be performed  (respectively, cases qDNS1 at 29 percent of the path and case qDNS at 50 percent of the path, Fig. \ref{scales}). This type of qDNS still captures the Ohmic dissipation fraction $f_{\Omega}$ well, and in any case better than LES (compare LES, qDNS1 and qDNS2 at 29 percent of the path in Fig. \ref{fohm}), but the stronger approximation implies a perturbed force balance structure at high harmonic degree (Fig. \ref{DNSforcebal}c) and an inaccurate determination of $\ell_\mathrm{CIA}$ (compare qDNS1 and qDNS2 in Fig. \ref{scales}). Finally, 
LES simulations (Fig. \ref{tql}a) tend to render a mix of CIA and VAC force balances at second order, with $\ell_\mathrm{CIA} \approx \ell_\mathrm{VAC}$ and both values being significantly decreased relative to DNS (Table \ref{bigtable2}, Fig. \ref{scales}). At path positions of 29 percent and beyond the ordering of these two length scales is even reversed compared to the DNS ordering (i.e. $\ell_\mathrm{VAC}<\ell_\mathrm{CIA}$, see violet symbols in Fig. \ref{scales}), implying that VAC has replaced CIA as the second-order force balance. This implies that the determination of $f_{\Omega}$ is inaccurate (Fig. \ref{fohm}), as already reported in A17.

\subsection{Spatial energy spectra and power-law ranges}
\begin{figure}
\centerline{\includegraphics[width=10cm]{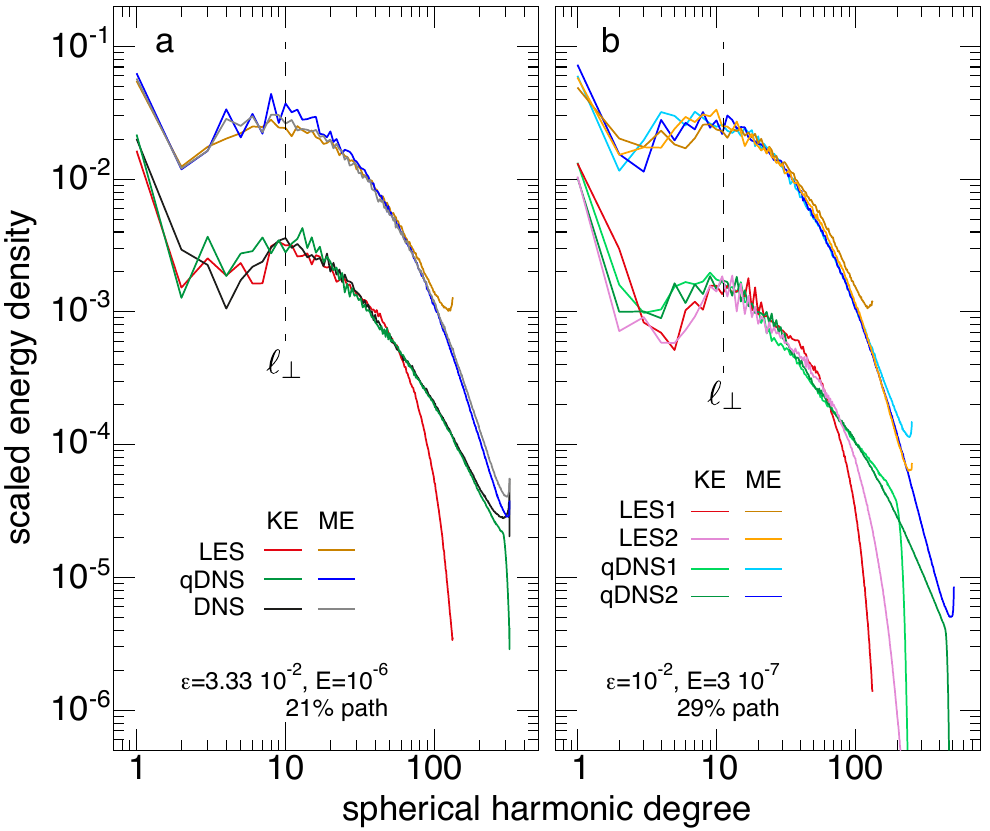}}
\caption{\label{spectratql} Instantanous kinetic (KE) and magnetic (ME) energy density spectra, as functions of the spherical harmonic degree $\ell$, in DNS, qDNS and LES simulations performed at 21 percent (a) and 29 percent (b) of the parameter space path (Ekman numbers $E=\te{-6}$ and $E=3~\te{-7}$). Spectra are normalised such that the total magnetic energy is one, illustrating the increasing dominance of the magnetic energy as we progress along the parameter space path.}
\end{figure}

Differences between upsizing strategies can be further assessed by examining spatial spectra of the kinetic and magnetic energy densities (Fig \ref{spectratql}). Here again, a representation using the spherical harmonic degree $\ell$ is used, but similar results can be obtained by using the spherical harmonic order instead. All simulations have broadly similar magnetic energy spectra throughout the spherical harmonic degree range, and also similar kinetic energy spectra up to spherical harmonic degree $\ell \approx 60$, underlining the relevance of LES at such large scales and the weak impact of upsizing on these scales. In particular, the peak of kinetic and magnetic energy spectra for $\ell>1$ occurs at $\ell=\ell_{\perp}$, confirming that $\ell_{\perp}$ is the dominant scale of velocity, density anomaly and magnetic field planforms (Figs. \ref{Vviz}-\ref{Bviz}). There is no evidence of a self-similar, power-law range in spectra of the magnetic energy density. This is expected given the modest level $Rm\approx 1000$ of magnetic turbulence (Table \ref{bigtable2}). In kinetic energy spectra of the LES simulations, a power-law decay range cannot be observed because it is obscured by hyperdiffusion kicking in at $\ell_{h}=30$ (red and pink curves in Fig. \ref{spectratql}a,b), but it appears gradually in qDNS and DNS simulations where the hyperdiffusive cut-off $\ell_{h}$ is increased close to $\ell_\mathrm{max}$ or removed altogether. It is again advisable to use a sufficiently resolved qDNS simulation (dark green curves in Fig. \ref{spectratql}a,b) for best accuracy (compare the dark green and black curves in Fig. \ref{spectratql}a). In particular, choosing $\ell_{h}$ above $\ell_{\Omega}$, but close to or below $\ell_\mathrm{CIA}$ causes energy stacking effects where the tail of the spectrum overshoots the true value (compare light green and dark green curves in Fig. \ref{spectratql}b). This in turn causes the buoyancy and Coriolis force lines to overshoot their true value in the force balance diagrams (Fig. \ref{DNSforcebal}c).

\begin{figure}
\centerline{\includegraphics[width=8cm]{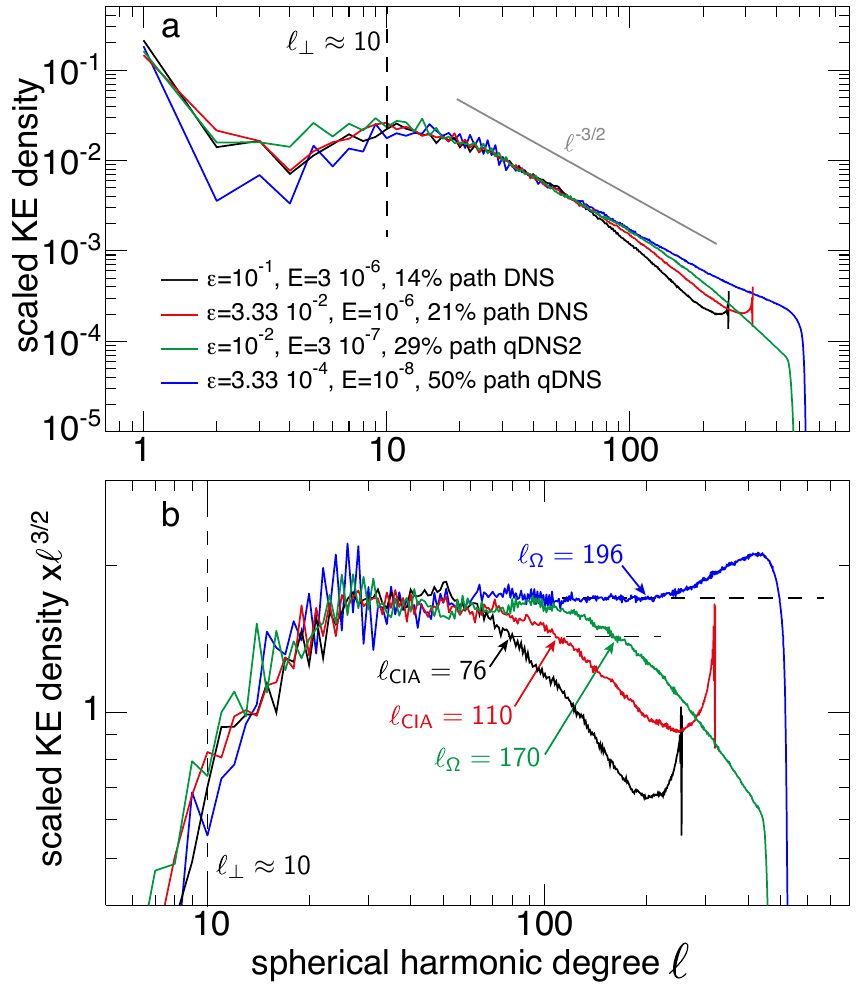}}
\caption{\label{evolpath} a: Instantaneous kinetic energy density spectra in DNS and and best-resolved qDNS simulations along the parameter space path, as functions of the spherical harmonic degree $\ell$. Spectra are normalised in each case relatively to the total kinetic energy. b: kinetic energy density spectra compensated with $\ell^{3/2}$. The horizontal dashed line in (b) represents 85 percent of the flat level of compensated kinetic energy density.}
\end{figure}

Plotting together uncompensated and compensated kinetic energy spectra (Fig. \ref{evolpath}a,b) of our best-resolved qDNS and DNS simulations along the path further reveals that the simulations are supportive of a gradually broadening range with power-law decay $\ell^{-3/2}$ starting from the large scale $\ell_\perp$. Less steep decay ranges have been previously observed in strongly scale-separated dynamo simulations \citep{Schaeffer2017,Sheyko2018}, with spectral indices from $-1$ to $-4/3$, and evidence of dependence of spectral index with the simulation forcing has also been reported. The steeper decay observed here occurs in a context of even stronger forcing (stronger $Re$ and $Rm$) than in these previous simulations and approaches closer to the index $-2$ predicted from vorticity equivalence (section \ref{powerlaw}). In our three simulations where resolution is high enough to avoid energy stacking at the spectrum tail (black, red and green curves in Fig. \ref{evolpath}), the extent of the power-law range is measured through the harmonic degree at which the compensated spectra of Fig. \ref{evolpath}b fall below 85 percent of the plateau energy. We then find that the corresponding harmonic degrees match the lowest of the two scales $\ell_\mathrm{CIA}, \ell_\Omega$ (see their relative position in Fig. \ref{scales}). This is theoretically expected because vorticity equivalence should be enforced down to the scale of magnetic dissipation, unless the MAC balance is perturbed at small scales by the second-order CIA balance. In our qDNS simulation at 50 percent of the parameter space path (blue curve in Fig. \ref{evolpath}) the same analysis cannot be carried out because of energy stacking at the spectrum tail, but we note that stacking again starts at $\ell_{\Omega}$ which in this case is significantly lower than $\ell_\mathrm{CIA}$ (Figs. \ref{DNSforcebal}c, \ref{scales}). Power transfers from injection at $\ell_{\perp}$ down to Ohmic dissipation at $\ell_{\Omega}$ are well controlled in this simulation (hence the flat compensated spectral range), but the transfer to even smaller scales of the residual power (3 percent in this case, Fig. \ref{fohm}) for viscous dissipation via inertial forces is improperly rendered (because the true value of $\ell_\mathrm{CIA}$ is not resolved), thereby causing the stacking of energy. Though this stacking appears significant in the compensated spectra (Fig. \ref{evolpath}b) it should not be overemphasized as it represents a small fraction of the kinetic energy (Fig. \ref{evolpath}a). Stacking is also commonly observed in fully-resolved DNS (see the tails of the black and red curves in Fig. \ref{evolpath}a,b), and in previously published simulations with moderate resolution \citep[see Fig. 3b of][]{Sheyko2018} without bearing too much consequence on the simulated dynamics.

\begin{figure}
\centerline{\includegraphics[width=10cm]{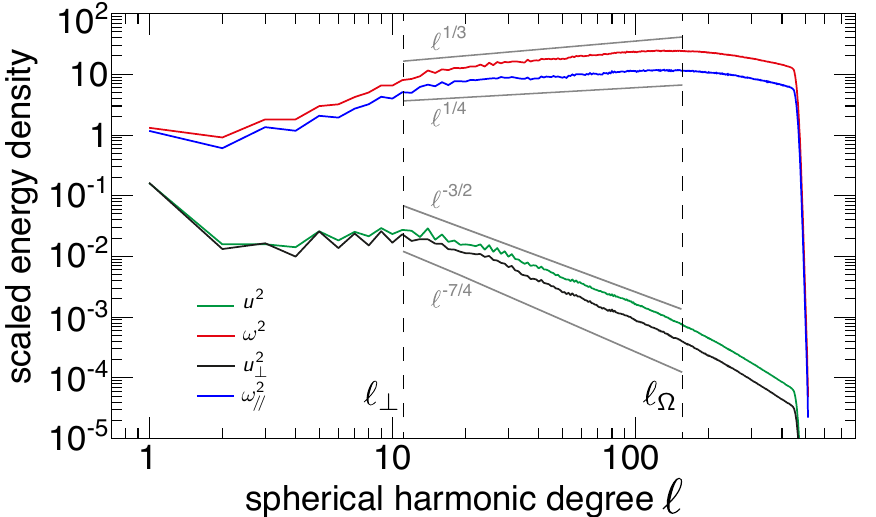}}
\caption{\label{UOm} Instantaneous energy density spectra of the enstrophy $\omega^{2}$, kinetic energy $u^{2}$, axial enstrophy $\omega_{\parallelsum}^{2}$ and kinetic energy of flow perpendicular to the rotation axis $u_{\perp}^{2}$, in case qDNS2 at 29 percent of the parameter space path (Ekman number $E=3~\te{-7}$). Spectra are normalised relatively to the total kinetic energy.}
\end{figure}

In order to further establish the link between Davidson's vorticity equivalence (equation \ref{vorteq}) and the power-law range observed in Figs. \ref{spectratql},\ref{evolpath}, spatial spectra of enstrophy $\omega^{2}$ are plotted together with kinetic energy spectra in Fig. \ref{UOm} for case qDNS2 at 29 percent of the parameter space path. At scales between the MAC scale $\ell_{\perp}$ and the magnetic dissipation scale $\ell_{\Omega}$, the enstrophy $\omega^{2}$ presents a weakly increasing power-law range $\omega^{2} \propto \ell^{1/3}$. Given that scale separation covers a decade in this example, this implies that the amplitude of vorticity varies by less than a factor 2 between the large and small scales, consistent with the idea of a flat spectrum underlain by vorticity equivalence (equation \ref{omegalaw}). The vorticity spectrum peaks at $\ell=\ell_{\Omega}$, showing that the dominant scale of the vorticity planforms observed in Fig. \ref{Vviz}f is indeed the scale of magnetic dissipation (the same analysis carried out in the case of the LES case of Fig. \ref{Vviz}e reveals that $\ell_\mathrm{VAC}$ is the dominant scale). Note that a length scale defined as $\delta(\ell)=u(\ell)/\omega(\ell) \sim D\ell^{-11/12}$ does not exactly vary like $\ell^{-1}$, underlining weak but measurable effects of the anisotropic flow structure (Fig. \ref{Vviz}b,d). Computing the spectra of axial enstrophy $\omega_{\parallelsum}^{2}$ removes the influence of the long length scale parallel to the rotation axis, and reveals a better respected vorticity equivalence  $\omega_{\parallelsum}^{2} \sim \ell^{1/4}$ in the range $[\ell_{\perp},\ell_{\Omega}]$. Likewise, the kinetic energy spectrum of flow perpendicular to the rotation axis follows $u_{\perp}^{2} \propto \ell^{-7/4}$, better approaching the spectral index $-2$ (equation \ref{ulaw}), and the associated perpendicular length scale $\delta_{\perp}(\ell)=u_\perp(\ell)/\omega_{\parallelsum}(\ell)$ now indeed varies like $\ell^{-1}$, confirming the relevance of spherical harmonic degrees to analyse length scales.

\subsection{Scaling properties}
\begin{figure}
\centerline{\includegraphics[width=6.8cm]{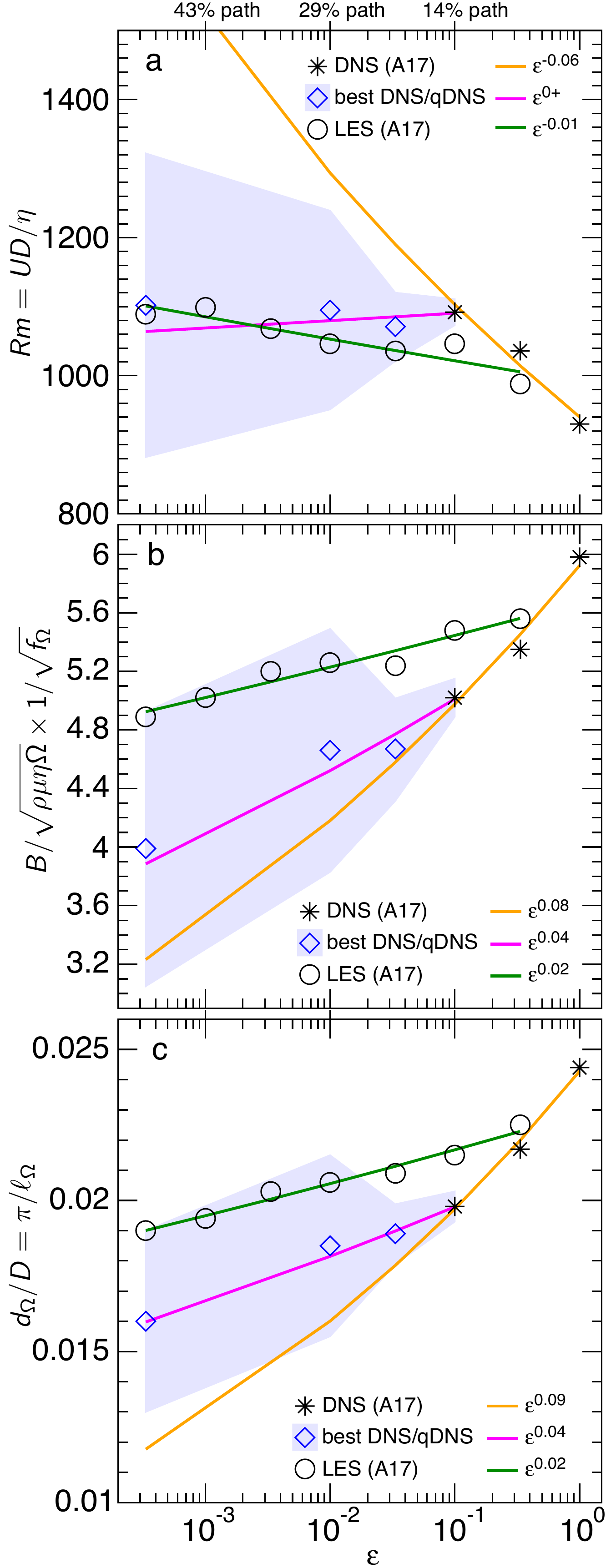}}
\caption{\label{scalings}Evolution of the magnetic Reynolds number $Rm$ (a), magnetic field amplitude $B/\sqrt{\rho\mu\eta\Omega}$ corrected as in A17 with the inverse square root of the ohmic fraction $f_\Omega$ (b), and magnetic dissipation length scale $d_{\Omega}$ (c) with the path parameter $\epsilon$ (Earth's core conditions are towards the left on these graphs). Presented are the set of LES from A17 (black circles), the set of DNS from A17 (black stars) and the best-resolved qDNS and DNS from this work (blue diamonds), with the blue shaded regions representing the uncertainty range associated to short-term time averaging of these runs (see text). Power laws obtained by least-squares fitting of the A17 results are also reported (orange and green lines for DNS and LES, respectively). The power-law fit obtained from DNS simulations (orange line) closely follows the predictions from the QG-MAC theory \citep[][A17]{Davidson2013}. For the new DNS and qDNS simulations, best-fitting power laws (purple lines) are obtained by using weighted least-squares taking the uncertainty range into account.}
\end{figure}

The outputs from our upsized qDNS and DNS simulations at highest resolution are expected to approach the asymptotic QG-MAC scaling theory of \cite{Davidson2013}, and to depart from the scale-independent path theory outlined in A17 and verified on LES simulations (see section \ref{asympt}). Fig. \ref{scales} indeed shows that $\ell_{\perp}$ in qDNS and DNS simulations tends to be slightly higher than in LES simulations, and to slightly increase along the path, both results being consistent with these expectations. However these results are not fully systematic, presumably because of our evaluation of $\ell_{\perp}$ on snapshots and the associated uncertainty, which while being small (about 10 percent) remains on the order of the observed variations. Furthermore, from equation (\ref{dperpscal}) a doubling of $\ell_{\perp}=\pi/d_{\perp}$ is predicted between 14 percent and 50 percent of the path, but clearly not observed. \rev{The departure of our results for $\ell_{\perp}$ from the QG-MAC prediction simply reflects the fact that the scale separation $d_{\perp}/d_{\Omega}$ in our simulations actually increases along the path, while from equations (\ref{dperpscal},\ref{dohmscal}) it is (rather counter-intuitively) expected to decrease in the QG-MAC theory.}

To refine this analysis, we further examine in Fig. \ref{scalings} the evolution along the path of the dimensionless velocity $Rm$, magnetic field $B/\sqrt{\rho\mu\eta\Omega}$ (corrected as in A17 with the inverse square root of $f_\Omega$ to account for incompletely Ohmic dissipation at the start of path), and magnetic dissipation length scale $d_{\Omega}/D$. To evaluate the uncertainties that are due to the shortness of DNS and qDNS runs, our longest LES simulation at 14 percent of the path (Table \ref{bigtable1}) is used as a reference. Error bars for other cases are then computed by evaluating in this reference case the maximum deviation of outputs averaged over the short duration of these cases versus the long time average of the reference. Power laws of DNS and qDNS are then obtained from the numerical data by using least-squares weighted with the resulting uncertainties. The magnetic Reynolds number of upsized simulations (Fig. \ref{scalings}a) remains about constant along the path, with a scaling $Rm \sim \epsilon^{0.004\pm 0.04}$. This result is close to the prediction from the path theory (constant $Rm$, A17), and deviates from the asymptotic QG-MAC prediction $Rm\sim \epsilon^{-1/18}\rev{\sim \epsilon^{-0.056}}$ (equation \ref{Uscal}, yellow line in Fig. \ref{scalings}a) in a significant manner even with uncertainties taken into account. The magnetic field of upsized simulations (Fig. \ref{scalings}b) follows the $f_{\Omega}$-corrected scaling $B/\sqrt{f_{\Omega}\rho\mu\eta\Omega} \sim \epsilon^{0.04\pm 0.06}$ and uncorrected scaling $B/\sqrt{\rho\mu\eta\Omega} \sim \epsilon^{0.02\pm 0.06}$, not fundamentally deviating from the constant prediction of the path theory (A17) but this time also marginally compatible with the QG-MAC prediction $B/\sqrt{\rho\mu\eta\Omega} \sim \epsilon^{1/12}\rev{\sim \epsilon^{0.083}}$ (equation \ref{Bscal}). Similar results are found for the magnetic dissipation length scale (Fig. \ref{scalings}c) that follows $d_{\Omega}/D \sim \epsilon^{0.04\pm 0.04}$, marginally compatible with the constant prediction from the path theory and the QG-MAC prediction $d_{\Omega}/D \sim \epsilon^{1/12}$ (equation \ref{dohmscal}). In summary, output diagnostics from the upsized qDNS and DNS simulations indeed better approach the QG-MAC theory of \cite{Davidson2013} than their LES counterparts, but fail to fully adhere to this theory, \rev{with the most important deviations being observed on the diagnostic for flow velocity}.

\section{\label{discu}Discussion}
By increasing the resolution and relaxing the hyperdiffusive approximation of the large-eddy simulations presented in A17, we have presented high-resolution numerical simulations of the geodynamo that approach closer to the rapidly rotating and turbulent conditions of Earth's core than earlier extreme simulations \citep{Yadav2016PNAS,Schaeffer2017,Sheyko2018}. The high-resolution models could be integrated only for a short time, but they are still relevant as they reach dynamical saturation after transients lasting only a fraction of an overturn time (Fig. \ref{transients}). This rapid equilibration owes to the fact that the large-eddy simulations that they originate from are already in the correct large-scale force balance (Fig. \ref{tql}). In that sense, large-eddy simulation followed by upsizing is an highly efficient strategy to reach extreme conditions at a fraction of the computer cost that would be involved in equilibrating the simulation from scratch. By reaching extremely low levels of the fluid viscosity in much of the harmonic degree range (up to five orders of magnitude below those of the leading forces, Fig. \ref{DNSforcebal}),  the upsized simulations also reach a state where the injected convective power is almost entirely dissipated by Ohmic losses (Fig. \ref{fohm}). Combined with the previous results from A17, the new high-resolution simulations provide insight into force balance, scale separation and turbulence in Earth's core.

\subsection{QG-MAC force balance in Earth's core}

A quasi-geostrophic equilibrium between Coriolis and pressure forces (Fig. \ref{DNSforcebal}) has been confirmed as the leading-order force balance enforced at large scales as we progress towards Earth's core conditions with high-resolution simulations. Though the magnetic force also reaches leading order at smaller scales \citep[as previously advocated by][]{Aurnou2017}, its contribution is essentially a magnetic pressure because of dynamic alignment of vorticity filaments and magnetic field lines (Fig. \ref{align}), implying that from a dynamical standpoint the whole length scale range in fact remains close to quasi-geostrophic. These results further rule out the possibility that Earth's core is in a system-scale magnetostrophic state, and should encourage further theoretical developments centered on quasi-geostrophy \citep{Calkins2015,Calkins2018}. At the next order, the magneto-Archimedes-Coriolis force balance already found in \cite{Yadav2016PNAS,Aubert2017} has also been confirmed in high-resolution simulations. 

\subsection{Scale separation in Earth's core}
In the vicinity of the large scale $\ell_{\perp}$ of the MAC balance, the upsized simulations preserve the large-scale features and axial invariance of the large-eddy simulations that they originate from. They however reveal an important scale separation in their velocity, vorticity and density anomaly fields (Figs. \ref{Vviz},\ref{Cviz}) while the magnetic field remains at large scales (Fig. \ref{Bviz}). The most interesting features revealed by upsizing are vorticity filaments that are most prominent at the small scale $\ell_{\Omega}$ of magnetic dissipation (Fig. \ref{UOm}). These exist at the edges of the large-scale, sheet-like convective upwelling and induce ramifications of density anomaly plumes. The filaments align with the large-scale magnetic field lines (Fig. \ref{align}), thereby minimising the non-pressure part of the associated Lorentz force at small scales (Fig. \ref{DNSforcebal}). The distribution of velocity and vorticity across scales has been analysed in the spectral space by using a decomposition of energy into spherical harmonic degrees. This approach is justified because at the strong forcing of our simulations, the effects of spatial heterogeneity and anisotropy are both reduced (Figs. \ref{Vviz},\ref{Cviz},\ref{UOm}) relative to simulations at weaker forcing \citep{Schaeffer2017,Sheyko2018}. The spectral analysis (Figs. \ref{evolpath},\ref{UOm}) has partly confirmed the principle of vorticity equivalence from the QG-MAC theory \citep{Davidson2013}, with approximately flat enstrophy spectra between degree $\ell_{\perp}$ (large scale $d_{\perp}$) at which  the magnetic field draws its power from convection and degree $\ell_{\Omega}$ (small scale $d_{\Omega}$) where this power is dissipated. High-resolution simulations feature a $\ell^{-3/2}$ power-law decay in the range $[\ell_{\perp},\ell_{\Omega}]$, approaching closer to the spectral index $-2$ predicted from vorticity equivalence than earlier simulations at lower forcing \citep{Schaeffer2017,Sheyko2018}. On a side note, \cite{Schaeffer2017} mention that with a kinetic energy density power-law decay not much steeper than $\ell^{-1}$, integral length scales built by weighting each harmonic degree $\ell$ with this density \citep{ChristensenAubert2006,King2013} become ill-posed. The present high-resolution simulations confirm this view and suggest that such measures should be abandoned altogether as they do not match any of the physically relevant crossover length scales identified here.

From the evolution of key length scales along the parameter space path (Figs. \ref{scales},\ref{scalings}), at the end of the parameter space $\epsilon=\te{-7}$ (i.e. at Earth's core conditions) we predict large-scale features at scale $\ell_{\perp}\approx 10$ and small scale features at $\ell_{\Omega}\approx 300$ (respectively $d_{\perp} =\pi D / \ell_{\perp} \approx 700 \ut{km}$ and $d_{\Omega}=\pi D / \ell_{\Omega} \approx 20 \ut{km}$), hence a significant but not extremely large level of scale separation. This implies that the large-scale structures currently imaged by core flow modelling \citep[see][for a review]{Holme2015} are representative of the large scales of the geodynamo, and that numerical dynamo simulations can provide robust a-priori constraints on the spectral decay of the flow in this inverse problem \citep[e.g.][]{Aubert2013a,Aubert2014,Aubert2015, Fournier2015}. 

\subsection{Optimal strategies for large-scale approximation of Earth's core dynamics}
In our system, the convective power $p$ is essentially carried by magnetic nonlinearities associated to the Lorentz force from the injection scale $\ell_{\perp}$ to the Ohmic dissipation scale $\ell_{\Omega}$ (section \ref{QGMAC}). To further render the transfer of the residual power $(1-f_{\Omega})p$ to even smaller viscous dissipation scales through hydrodynamic nonlinearities, one needs to resolve at least the Coriolis-Inertia-Archimedes scale $\ell_\mathrm{CIA}$ (Figs. \ref{fohm}, \ref{tql}). From the simulation cases where $\ell_\mathrm{CIA}$ is correctly resolved (Fig. \ref{scales}) we can infer the dependency $\ell_\mathrm{CIA} \sim \epsilon^{-0.4}$, leading at Earth's core conditions to $\ell_\mathrm{CIA} \approx 20000$, or the equivalent length scale $\pi D/ \ell_\mathrm{CIA} \approx 400 \ut{m}$. Though the need to resolve such a small scale obviously complexifies the numerical simulation problem, we note that this residual part of power $(1-f_{\Omega})p$ becomes vanishingly small in our simulations and therefore at Earth's core conditions (Fig. \ref{fohm}), thereby motivating further research towards approximated simulations of moderate maximal resolution $\ell_\mathrm{max} \approx  \ell_{\Omega} \ll \ell_\mathrm{CIA}$. Introducing quasi-direct numerical simulations where the hypderdiffusive cut-off $\ell_{h}$ has been set to a value larger than $\ell_{\Omega}$, but smaller than $\ell_\mathrm{CIA}$, we have for instance obtained an acceptable trade-off between computational cost and accuracy, with an accurate value of the Ohmic dissipation fraction $f_{\Omega}$ (Fig. \ref{fohm}) but some stacking towards the kinetic energy spectrum tail (Figs. \ref{spectratql},\ref{evolpath}) and some degradation of the small-scale force balance (Fig. \ref{DNSforcebal}). The upsized results have also confirmed the quality of earlier large-eddy simulations from A17, which capture the large-scale planforms (Figs. \ref{Vviz}-\ref{Bviz}), the leading order and first order force balances (Fig. \ref{tql}) and dynamical equilibria such as the enforcement of the Taylor constraint (Table \ref{bigtable2}) to an excellent level of accuracy, but fail to render the partition of dissipation between Ohmic and viscous losses (Fig. \ref{fohm}) because they introduce a viscous-Archimedes-Coriolis force balance at second order. 

\subsection{Rotating magnetohydrodynamic turbulence in the spherical shell geometry}
It is interesting to compare the turbulence properties observed here to the non-magnetic, but rotating case. In classical two-dimensional and quasi-geo\-strophic turbulence \citep[e.g.][]{Nataf2015}, the enstrophy cascades to smaller scales through vortex stretching. In the present high-resolution simulations, vortex stretching is strongly constrained by dynamic alignment of vortex filaments with magnetic field lines (Fig. \ref{align}). This may explain why the velocity spectrum decay $\ell^{-3/2}$ (Figs. \ref{evolpath}, \ref{UOm}) is less steep than the decay $\ell^{-5/3}$ previously reported in a strongly forced simulation of nonmagnetic and rotating convection \citep{Sheyko2018}, and much less steep than the decay $m^{-5}$ obtained along spherical harmonic orders in simulations of quasi-geostrophic Rossby wave turbulence \citep{Schaeffer2005}. 

Examining our results within the framework of non-rotating, magnetohydrodynamic turbulence, we first note that our simulations are in the strong turbulence regime where magnetic field lines are bent by velocity fluctuations. Cartesian simulations of strong MHD turbulence commonly feature energy density spectra decaying like $k_{\perp}^{-3/2}$, where $k_{\perp}$ is the field-perpendicular wavenumber, and scale-dependent dynamic alignment of vorticity structures along field lines is essential to reconcile this spectral decay with existing theories \citep[see e.g.][]{Tobias2012}. It is interesting to note that dynamic alignment in the physical space is also clearly observed here, and that Alfvén waves, the energy carriers of strong magnetohydrodynamic turbulence, have been highlighted as important energy carriers in our simulations, both the axisymmetric and non-axisymmetric levels \citep{Aubert2018}. \rev{Interpreting the common spectral decay index -3/2 of our results and the strong MHD turbulence theory is however subject to caution, as the relationship between the cartesian wavenumber $k_{\perp}$ and the spherical harmonic degree $\ell$ in volume-averaged spectra is not straightforward \citep{Schaeffer2017}.}

\subsection{Towards an ultimate asymptotic scaling theory}
The outputs from the best-resolved upsized simulations approach the scaling predictions of the QG-MAC theory \citep{Davidson2013} better than their large-eddy counterparts, but do not fully adhere to this theory. \rev{The deviations are clearest for the length scale $d_{\perp}$ (Fig. \ref{scales}) and the flow velocity or magnetic Reynolds number $Rm$  (Fig. \ref{scalings}). They} already appear in a DNS carried out at 21 percent of the path, suggesting that this is not the result of residual hyperdiffusivity effects. \rev{It remains possible that the magnetic equilibration of the upsized simulations takes place on a time scale that is still longer than the short integration times that could be achieved here, in which case the error bars specified of Fig. \ref{scalings} would be underestimated. If such is the case however, the DNS at 21 percent of the path remains puzzling as its magnetic field amplitude appears to almost match the QG-MAC prediction while the velocity field amplitude strongly deviates from this theory.} Deviations from the QG-MAC theory may be needed somehow, \rev{because we have seen that this theory fails to predict an increase of scale separation along the path as Earth's core conditions are approached, in apparent contradiction with the present results (Fig. \ref{evolpath}) and previous high-resolution numerical simulations \citep{Sakuraba2009,Yadav2016PNAS,Schaeffer2017,Sheyko2018}. It has also} been shown in A17 that the predictions from the scale-invariant path theory match the amplitude of Earth's core magnetic and velocity fields strikingly well, and better than the QG-MAC predictions. In this respect, it is possible that the vorticity equivalence (spectral index -2 in the kinetic energy spectrum) is incompatible with dynamic alignment of vorticity and magnetic field lines (spectral index -3/2). When stacked at advancing positions along the parameter space path (Fig. \ref{evolpath}), kinetic energy spectra do indeed appear to converge towards index -3/2 and there is no evidence that the index could evolve towards -2 upon further progress along the path. If index -3/2 is the asymptotic value, then the vorticity equivalence constant $c_{\omega}$ in equation (\ref{vorteq}) may have some residual dependence on the path parameter $\epsilon$, while still remaining of order one at the end of path. This indicates that some refinements may still be needed for an ultimate scaling theory of Earth's core dynamo. Further work should account for the gradual enforcement of structuring constraints exerted by the magnetic field on the flow as we progress towards Earth's core conditions.

\section*{Acknowledgements}
JA wishes to thank Thomas Gastine for insightful discussions and suggestions, and acknowledges support from the Fondation Simone et Cino Del Duca of Institut de France (2017 Research Grant). Numerical computations were performed at S-CAPAD, IPGP and using HPC resources from GENCI-IDRIS, GENCI-TGCC and GENCI-CINES (Grants A0020402122 and A0040402122). This is IPGP contribution 4022.

\bibliographystyle{gji}
\bibliography{Biblio}

\end{document}